\documentclass[tc]{copernicus}

\usepackage{tabularx}
\usepackage{stmaryrd}
\usepackage{graphicx}
\usepackage{natbib}
\usepackage{placeins}

\newcommand{\vv}{\mathbf{v}}
\newcommand{\DD}{\underline{\bs{D}}}
\newcommand{\beq}{\begin{equation}}
\newcommand{\eeq}{\end{equation}}
\newcommand{\eg}{\textit{e.g. }}
\newcommand{\ie}{\textit{i.e. }}
\newcommand{\vu}{\mathbf{u}}
\newcommand{\bs}{\boldsymbol}
\newcommand{\bk}{\mathbf{k}}
\newcommand{\dx}{~d\bs{x}}

\newcommand{\mM}{\mathcal{M}}

\newcommand{\mK}{\mathcal{K}}
\newcommand{\mY}{\mathcal{Y}}
\newcommand{\bega}{\begin{array}}
\newcommand{\ena}{\end{array}}
\DeclareMathOperator{\di}{\boldsymbol{div}} 
\DeclareMathOperator{\Id}{\boldsymbol{Id}}
\usepackage{accents}

\runningauthor{N. Martin and J. Monnier}
\runningtitle{Adjoint accuracy for the full-Stokes ice flow model}

\begin{document}

\title{Adjoint accuracy for the full-Stokes ice flow model:
  limits to the transmission of basal friction variability to the surface}

\author[1,2]{Nathan Martin}
\author[1,2]{Jérôme Monnier}
\affil[1]{INSA, 135 Avenue de Rangueil, 31077, Toulouse Cedex 4, France}
\affil[2]{IMT, 118, route de Narbonne, F-31062 Toulouse Cedex 9 }
\correspondence{N. Martin\\ (nmartin@insa-toulouse.fr)}

\maketitle

\label{chap5}

\begin{abstract}
  This  work focuses  on the  numerical  assessment of  the accuracy  of an  adjoint-based
  gradient   in  the   perspective  of   variational  data   assimilation   and  parameter
  identification in  glaciology.  Using noisy synthetic  data, we quantify  the ability to
  identify the friction coefficient for such  methods with a non-linear friction law.  The
  exact  adjoint  problem is  solved,  based  on second  order  numerical  schemes, and  a
  comparison with  the so called ``self-adjoint'' approximation,  neglecting the viscosity
  dependency to the velocity (leading to  an incorrect gradient), common in glaciology, is
  carried out.  For data with a noise  of $1\%$, a lower bound of identifiable wavelengths
  of $10$ ice thicknesses in the friction coefficient is established, when using the exact
  adjoint method, while the ``self-adjoint'' method is limited, even for lower noise, to a
  minimum of  $20$ ice  thicknesses wavelengths.  The  second order exact  gradient method
  therefore provides robustness and  reliability for the parameter identification process.
  In other respect, the derivation  of the adjoint model using algorithmic differentiation
  leads  to  formulate a  generalization  of  the  ``self-adjoint'' approximation  towards
  an \textit{incomplete adjoint method}, adjustable in precision and computational burden.
\end{abstract}

\introduction

The main available observations of the cryosphere are generally obtained from remote-sensed 
techniques and are thus essentially surface observations. However, ice dynamics
is known to be highly sensitive to the state of the bed (and
therefore to how the bed is modeled, see \eg \citet{paterson2010}),
not to the surface which is more easily observable. The friction coefficient is consequently a critical parameter in terms of controlling
ice flows. This raises questions about, on one hand, whether the
surface can provide the necessary information about basal conditions
and, on the other hand, whether inverse methods can adequately recover
this information.\\

Many authors have address the first question by investigatin how 
bedrock topography affects the surface. \citet{balise1985} conducted
one of the earliest studies concerning the transmission
of fluctuations in basal  slip to the surface for a
Newtonian fluid, using  perturbation methods. The non-local aspect  of the transmission of
the variations of the friction coefficient at surface is established by \citet{raymond1996}
where it is dependent  upon  the  slip  ratio  (ratio  between mean  sliding  velocities  and  mean  ice
deformation velocities). These queries  are extended in \citet{gudmundsson2003}, still
under the Newtonian hypothesis using perturbation methods.  In these
studies, one of the main conclusions is that
 the  transmission of  basal  variability at the surface increases with increased
sliding.

The question of the representability  of the friction coefficient through surface velocity
observations  (horizontal   and  vertical)   using  an  inverse   method  is   studied  by
\citet{gudmundsson2008}.  The  method, based on a  Bayesian approach, is used  to study the
effect of density and  quality of surface velocity data on the  estimation of the friction
coefficient for  a Newtonian fluid  and a  linear sliding law. In the
reconstruction of small amplitude variations of the friction
coefficient, a wavelength limit of around $50$ times the ice thickness
is found. A similar method in the case of a non-Newtonian fluid
and a non-linear sliding law is developed in \citet{raymond2009}.

In other respects,  the identification method based on  \citet{macayeal1993} and widely used
(see \eg  \citet{larour2005,joughin2004,morlighem2010}) makes the assumption  that viscosity
is independent of the velocity, and a limited attention has been paid to the quality of the
resulting estimations  in terms  of spatial variability  of the friction  coefficient (see
\citet{gudmundsson2008}). Comparisons with the ``self-adjoint'' method and the
use of an  exact adjoint are made by \citet{goldberg2011}, based  on a vertically integrated
approximation and by \cite{morlighem2013} based on the higher-order
model.   Limitations for  the minimizing  process are  highlighted by
\cite{goldberg2011} when  using the ``self-adjoint'' method.  To  the
best of our  knowledge, the use of an  exact adjoint in a
glaciological context for
the full-Stokes  problem has been done only by  \citet{Petra2012}. A
comparison between their results and
the results of  \citet{gudmundsson2008} on an academic problem allowed  then to conclude that
the exact  adjoint is  able to recover  wavelengths in the friction  coefficient of
approximately $20$ times the ice thickness in the case of a linear sliding law.\\

The  purpose  of  this study  is  the  numerical  evaluation  of  the limitations  of  the
``self-adjoint'' method compared to  the method using the exact
adjoint solution, referred as the full adjoint method in what
hereafter. The ``self-adjoint'' approximation for the full-Stokes problem
is detailed in terms of equations and presented as  a limited  case of  the reverse  accumulation method  used to  compute the
adjoint when  obtained using source-to-source automatic differentiation.   From a strictly
numerical perspective, tests on the accuracy reached by the gradients for both
methods are performed, demonstrating an  important limitation for the gradient computed by
the ``self-adjoint'' method.  We then  study the identifiability, for a non-linear sliding
law,  of high frequencies  in the  friction coefficient  depending on  the level  of noise
considered on synthetic  data. The quality of the estimations provided  by both methods is
compared in the case of dense horizontal surface velocity observations for a quasi-uniform
flow and then for a realistic flow presenting an important spatial variability.  The
realistic case is then applied for less dense data.\\

\section{Forward and adjoint model}

In this  section, we briefly present  what shall be  referred to hereafter as  the forward
model and describe the derivation of the  adjoint model and the computation of the adjoint
state.

\subsection{Forward model}
\label{formod}
\begin{figure}
        \includegraphics[width=.5\textwidth]{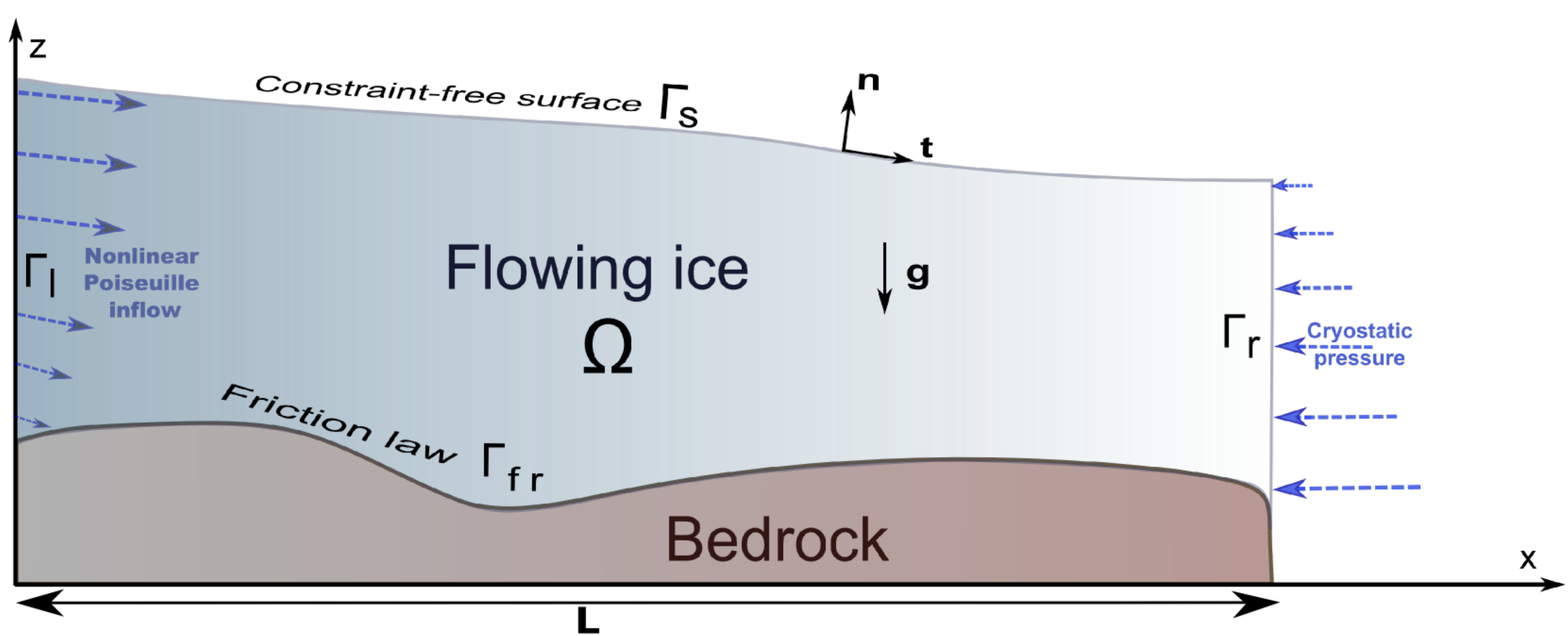}
        \caption{Geometry, boundary conditions and notations of the problem}
    \label{geom}
\end{figure}

The flow model  considered here is the bidimensional flowline power-law Stokes  model applied to a
gravity driven flow  (see \eg \citet{paterson2010}) and solved on  a given domain $\Omega$
of horizontal extent $L$ (see Figure \ref{geom}):
\begin{eqnarray} \hspace{1in} \mathbf{div}(\mathbf{u}) =& 0&~\mbox{ in }
\Omega, \label{debstokes}\\ -\mathbf{div}(2\eta(\mathbf{u})\underline{\bs{D}})+ \nabla p =&
\rho \mathbf{g}&~\mbox{ in } \Omega, \label{momstokes}\\
\eta(\vu) = \eta_0\|\underline{\bs{D}}\|_F^{\frac{1-n}{n}}.\label{viscol}
\end{eqnarray}
where   $\underline{\bs{\sigma}}=\eta(\mathbf{u})\underline{\bs{D}}   -  p\boldsymbol{Id}$
represents  the  Cauchy stress  tensor (with $\bs{Id}$ the
second-order two-dimension identity tensor),  $\eta(\mathbf{u})$  the  viscosity, $\eta_0$  the
consistency of the fluid, $n$ the power-law exponent, $\underline{\bs{D}}$ the strain rate
tensor, $\mathbf{u}=(u_x,u_z)$ the velocity field defined in the
Cartesian frame $(x,z)$,  $p$ the pressure field, $\rho$ the ice
density,  $\mathbf{g}$ the gravity  and $\|\underline{\bs{D}}\|^2_F  = \underline{\bs{D}}:
\underline{\bs{D}}$ the Frobenius matrix norm.\\

A Weertman-type sliding law is then prescribed at the bedrock boundary $\Gamma_{fr}$:
 \begin{eqnarray}\label{fric1}
|\mathbf{\sigma}_{nt}|^{m-1}\mathbf{\sigma}_{\mathbf{n}\mathbf{t}} =&\beta
\mathbf{u}\cdot\mathbf{t}&\mbox{ on } \Gamma_{fr},\\ \mathbf{u}\cdot\mathbf{n} =& 0
&~\mbox{ on } \Gamma_{fr}. \label{finstokes}
\end{eqnarray}
where $\beta=\beta(x)$ is a spatially variable parameter and 
where $(\bs{t},\bs{n})$, the tangent-normal pair of unit vectors, is such that:
\begin{equation} \underline{\bs{\sigma}} = (\underline{\bs{\sigma}} \cdot \bs{n})\bs{n} +
(\underline{\bs{\sigma}} \cdot \bs{t})\bs{t} \quad \eeq and: \beq
\underline{\bs{\sigma}} \cdot \bs{n} = \sigma_{nn}\bs{n} + \sigma_{n t}\bs{t} \quad,
\quad \underline{\bs{\sigma}} \cdot \bs{t} = \sigma_{t n}\bs{n} +
\sigma_{t t }\bs{t} \label{sigmant}\end{equation}

A  velocity profile corresponding  to the  solution of  the Stokes  problem for  a uniform
steady flow  of a parallel sided slab on an inclined bed with non-linear  friction defined by \eqref{fric1}  at the
bottom     is     prescribed     on     the     inflow     boundary.     This     solution
$\mathbf{u}=(u_{\overline{x}},u_{\overline{z}})$,   expressed    in   the   ``mean-slope''
reference frame $(\overline{x},\overline{z})$, is written (see \eg
\cite{martin2013}):
\begin{equation}
\label{permsol}
\begin{array}{c} u_{\overline{x}}(\overline{z})=\displaystyle{\frac{(-\rho \mathbf{g}
sin(\theta)h)^m}{\beta}} + \\ \displaystyle{\frac{1}{1+n}(2\eta_0)^{-n} (\rho \mathbf{g}
sin(\theta))^{n}(h^{1+n}-(H-\overline{z})^{1+n})}, \\ \end{array}
\end{equation}
\begin{equation} \displaystyle{u_{\overline{z}}=0},
\end{equation}
\begin{equation} \displaystyle{p(\overline{z})=\rho g cos(\theta)(H-\overline{z})}.
\end{equation}
with $\theta$ defining the  slope of the slab, $H$ the height  of the upper-surface and
$h$ the thickness.

A hydrostatic  pressure is considered on  the outflow.  All the  simulations are performed
with  an exponent  $m=3$ for  the sliding  law.  The domain is
discretized using triangular Taylor-Hood  finite elements and the
solution of  the  continuous forward problem is obtained using a
classical fixed point algorithm.  The geometry and notations of the problem are plotted in Figure \ref{geom}.\\

The  sensitivities  and  identifications  carried   out  in  this  work  use  adjoint-based
computation  and thus  require  the solution  of  the adjoint  problem  associated with  the
full-Stokes model. \\

All the computations are performed using the software DassFlow
(\cite{dassflow}). The fixed point algorithm
is used here as a typical iterative method for solving of the
full-Stokes problem but the assessments on the precision and
efficiency of the adjoint-based inverse problems should be valid for
any iterative algorithm. The details on the different approaches
used in DassFlow for the solution of the power-law Stokes problem can
be found in \cite{siam2013}.

\subsection{The basic principles of the adjoint model}

The output  of the forward  model is  represented by a  scalar valued function  $j$ called a
\textit{cost  function},  which depends on  the  parameters of  the  model  and represents  a
quantity to be minimized.  In presence of observations, part of the cost measures the
discrepancy (the misfit) between the computed state and an observed state (through any type of data).

The parameters of interest are  called \textit{control variables} and constitute a control
vector $\bs{k}$.  The  minimizing procedure operates on this control  vector to generate a
set  of  parameters which  allows  a  computed state  closer  to  the  observations to  be
obtained. In the following the control  vector includes only the friction
coefficient field $\beta(x)$. The corresponding optimal control problem can be written:
\begin{equation}
\underset{\bs{k}}{ Min }\hspace{0.4em}j(\bs{k})
\label{pb.control}
\end{equation}

This optimization problem  is solved numerically by a descent algorithm.  Thus, we need to
compute the gradient of the cost function.  This is done by introducing the adjoint model.\\ 

\subsection{Cost Function, Twin Experiments and Morozov's Discrepancy Principle} 
\label{morozovsec}
The cost function used for the identification is defined by:
\begin{equation}\label{costfunc} j(\beta;\gamma) = \frac{1}{2}\int_{\Gamma_s} \|u^{obs}_s(\beta_t) -
u_s(\beta)\|_2^2~~d\mathbf{x} + \gamma \mathcal{T}(\beta'),
\end{equation}

where the data $u^{obs}_s$ are synthetic horizontal surface
velocities obtained using a given friction coefficient $\beta_t$ and perturbed
with a random Gaussian noise of varying level $\delta$. The term $\mathcal{T}(\beta')$
 called Tikhonov's regularization controls the oscillations of the control
variable gradient $\beta'$. It is defined by:
\begin{equation}\label{costfunc2} \mathcal{T}(\beta') = \int_{[0,L]} \|\beta'
\|^2_2 ~ds.
\end{equation}

where $L$ is the length of the domain. The  parameter
$\gamma$ quantifies the  strength of the imposed  smoothness. This
term regularizes the functional to be minimized and introduces a bias
toward smoothly varying field.  The  tuning of these weights
can be achieved  from various considerations generally related to the  quality of the data
(or the noise level) and the degree of smoothness sought on the control variable. A
classical   approach,  referred to  as  the   Morozov's  discrepancy   principle   (see  \eg
\citet{vogel2002}), consists of choosing
$\gamma$ such that $j(\beta;\gamma)=j(\beta_t;0)$ \ie when the final
cost matches the noise level on the data. The methodology that
consists of using noisy synthetic data in order to retrieve a
set of reference parameters (here defined as $\beta_t$) a priori known
is called a twin experiment. The gradient of the cost function is given
by solving the adjoint problem and used by the algorithm to
compute at each iteration a new set of parameters in order to make the
cost $j$ decrease until convergence.\\

\subsection{Derivation of the adjoint model}
In order to efficiently compute all partial derivatives of a cost function $j(\bs{k})$
with respect  to the components of a  \textit{control vector} $\bs{k}$, we  introduce the adjoint
model (see \textit{e.g.} \citet{lions71}). 

In DassFlow software,  the adjoint model is obtained  by using algorithmic differentiation
of  the source  code (see  \citet{honnorat2007b,Honnorat2007,dassflow}).   This last  approach ensures  a
better consistency between  the computed cost function and its gradient, since
it is the computed  cost function that is differentiated.  A large  part of this extensive
task can be  automated using automatic differentiation (see  \citet{griewank1989}).  In the
case of DassFlow-Ice, the direct code is written in Fortran 95 and is derived using the
automatic differentiation tool Tapenade (see \citet{Hascoet2004}). The
linear solver used is MUMPS (\cite{amestoy2001}) and the
differentiation of the linear system solving process is achieved
using a ``bypass'' approach which considers the linear solver as an
unknown black-box (see appendix \ref{annexe}). This approach is
similar to the one used by \citet{goldberg2013}.\\

Let  $\mK$ be the  space of  control variables  and $\mY$  the space  of the  forward code
response.  In the present case, we have~:
$$
\bs{k}=(\beta) \mbox{ and } Y=(y,j)^T
$$

where $\beta$ is defined by  \eqref{fric1}.\\ 

Let us point out  that we include both the state and the cost  function in the response of
the   forward   code.   The   direct    code   can   be   represented   as   an   operator
$\mM\,:\mK\longrightarrow\mY$ such that:
$$
\mY = \mM(\mK) .
$$

The tangent model becomes $\frac{\partial\mM}{\partial\bs{k}}(\bs{k})\,:\mK\longrightarrow\mY$. 
As input variable, it takes a perturbation of the control vector $d\bs{k}\in\mK$, it then
gives the variation $dY\in\mY$ as output variable:
$$
    dY=\frac{\partial\mM}{\partial \bs{k}}(\bs{k})\cdot d\bs{k}\ .
$$

The adjoint model is defined as the adjoint operator of the tangent model. 
This can be represented as follows:
$$\left(\frac{\partial\mM}{\partial \bs{k}}(\bs{k})\right)^*\,:\mY'\longrightarrow\mK' .$$

It takes $dY^*\in\mY'$ as an input variable and provides the adjoint variable $d\bs{k}^*\in\mK'$ 
at output:

$$
    d\bs{k}^*=\left(\frac{\partial\mM}{\partial \bs{k}}(\bs{k})\right)^*\!\!\cdot dY^*\ .
$$

Now, let us  make the link between the adjoint code  and the gradient $\frac{dj}{d\bs{k}}$
we seek to compute.  By definition of the adjoint, we have~:
\begin{equation}
    \label{eq_def_programme_adjoint}
    \textstyle
    \Big\langle\left(\frac{\partial\mM}{\partial \bs{k}}\right)^*\!\!\cdot dY^*,\,d\bs{k}\Big\rangle_{\mK'\times\mK}=\
\Big\langle          dY^*,\,\left(\frac{\partial\mM}{\partial          \bs{k}}\right)\cdot
    d\bs{k}\Big\rangle_{\mY'\times\mY}\ .
\end{equation}
It reads, using the relations presented above:
\begin{equation}
    \label{eq_def_variables_adj}
    \textstyle
    \big\langle d\bs{k}^*,\,d\bs{k}\big\rangle_{\mK'\times\mK}=\ \big\langle dY^*,\,dY\big\rangle_{\mY'\times\mY}\ .
\end{equation}

If we set $dY^*=(0,1)^T$ and by denoting the perturbation vector 
$d\bs{k}=\left(\delta\beta \right)^T$, we 
obtain:

\begin{equation*}
\bega{lcr}
   & \textstyle
    \left\langle\left(\begin{array}{c}
        0 \\ 1
    \end{array}\right),\,
    \left(\begin{array}{c}
        dy^* \\ dj^*
    \end{array}\right)
    \right\rangle_{\mY' \times \mY}\  
    
  &  =\  \Bigg\langle\left(\begin{array}{c}
        \delta \beta^* 
    \end{array}\right),\left(\begin{array}{c}
        \delta \beta 
    \end{array}\right)\Bigg\rangle_{\mK'\times\mK} .
    \\
    \ena
\end{equation*}

Furthermore, we have by definition:
\beq
       dj\ =\  \frac{\partial j}{\partial \beta}(\bs{k})\cdot\delta \beta  .      
\eeq
Therefore, the adjoint variable $d\bs{k}^*$ (output of the adjoint code with $dY^*=(0,1)^T$) 
corresponds to the partial derivatives of the cost function $j$:
\beq\label{outputadj}
        \frac{\partial j}{\partial \beta}(\bs{k}) = \beta^* .
\eeq

A  single integration  of the  forward  model  followed by  a
single integration of the adjoint model allow  to compute all components
of the gradient of the cost function.\\ 

\begin{figure} \centering
        \includegraphics[scale=0.4]{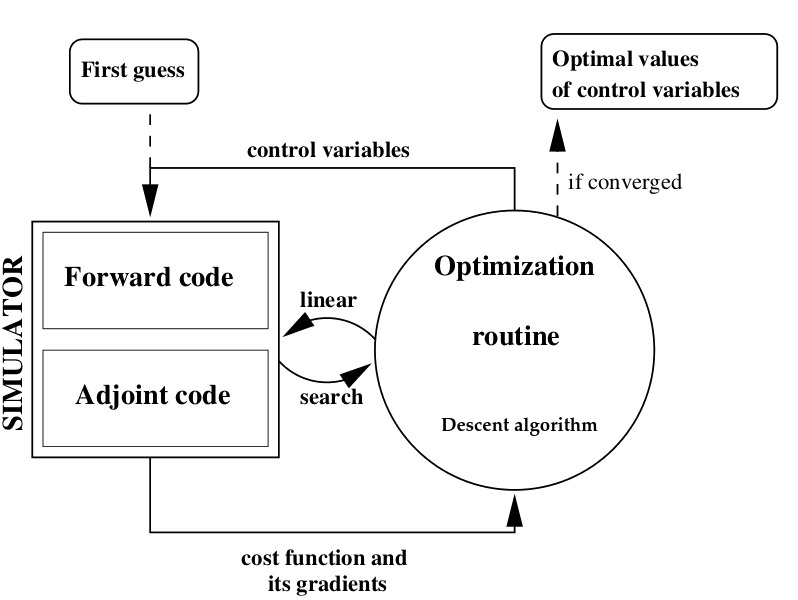}
        \caption{Principle of a 3D-Var type variational
          data assimilation algorithm.}
    \label{fig:daprincipe}
\end{figure}

The optimal control problem (\ref{pb.control})  is solved using a local descent algorithm,
more  precisely the L-BFGS  algorithm (a  quasi-Newton method),  implemented in  the M1QN3
routine (see \citet{Gilbert1989}).  Thus, these partial  derivatives are used as input to the
minimization algorithm  M1QN3. The  global optimization process  is represented  in Figure
\ref{fig:daprincipe}.\\

\subsection{The gradient test}\label{gradtestsec}
The gradient test is a classical adjoint code validation test and is
used hereafter in order to assess the precision of the
``self-adjoint'' approximation. The test aims to verify that
the partial derivatives of the cost function are correctly computed by comparing it with a
finite difference approximation (see \eg \citet{Honnorat2007} for the detailed test procedures).

Let us consider the following order two central finite difference approximation of the gradient:

\begin{equation}
\frac{j(\bk+\alpha\delta \bk) - j(\bk-\alpha\delta \bk)}{2\alpha} = \frac{\partial
  j}{\partial \bk}\cdot \delta \bk + O\left(\alpha^2 \delta \bk^3\right)
\label{central_FD}
\end{equation}
with $d\bs{k}=\alpha \delta \bs{k}$. This scheme leads us to define:
\begin{equation}\label{ialpha2}
    I_\alpha\ =\ \frac{j(\bk+\alpha\,\delta \bk)-j(\bk-\alpha\,\delta \bk)}{2\alpha\;\frac{\partial j}{\partial \bk}
(\bk)\cdot\delta \bk}\ .
\end{equation}

According to  \eqref{central_FD}, one must have: 
$\displaystyle\lim_{\alpha\rightarrow 0}\:I_\alpha=1$.\\
 The gradient test consists of verifying this property.\\

\section{``Self-adjoint'' approximation, full adjoint and reverse accumulation}

The  model considered here has been obtained  using algorithmic (or  automatic) differentiation of
the source code. Automatic differentiation of  a fixed point type iterative routine of the
form $y=\Phi(y,u)$ (such  as the solution of the non-linear Stokes  problem using a Picard
method) is carried out by reverse accumulation (see \citet{griewank1989,griewank1993}). The
reverse accumulation technique consists of building a computational graph for the function
evaluation where  the nodes  of the  graph represent every  value taken  by the
function. To every  node, an adjoint quantity containing  the gradient of the
function $\Phi$ with respect to the node, is associated.

The adjoint values are computed in reverse order. The final value of the gradient is given
by the sum  of the partial derivatives of  the function of the nodes  of the computational
graph. This  result is  a consequence of  the chain  rule. This process  \textit{a priori}
requires the storing of as  many states  of the system  as iterations  performed by  the forward
solver to reach the converged state.\\ 

It is shown by \citet{christianson1994} that,  in the case of a forward computation carried
out by  a fixed point method,  the adjoint quantity  also satisfies a fixed  point problem
whose rate  of convergence is  at least equal  to the rate  of convergence of  the forward
fixed point.   Based on  this result, it  is \textit{a  priori} necessary to  retain every
iteration of the  forward run to evaluate the gradient. In  practice, as further detailed
in section \ref{troncapp}, the number of  reverse iterations required to obtain an adjoint
state  with the  same precision  of the  forward state  can be  adjusted depending  on the
convergence speed of the direct construction.\\

\subsection{The ``self-adjoint'' approximation}

The ``self-adjoint'' method  in glaciology, applied to the shelfy-stream approximation, has
been proposed by \citet{macayeal1993}.  The approximation consists of deriving the adjoint
equation  system without  taking into  account the  explicit dependency  of  the viscosity
$\eta$ on  the velocity  field $\vu$. Let us recall that  the terminology
\textit{self-adjoint} only makes  sense in the Newtonian case ($n=1$).  It is important to
precise that the gradient resulting from this procedure is therefore an incorrect gradient.\\

For the full-Stokes case, the adjoint system considered under this
approximation is the adjoint associated to the forward problem
\eqref{debstokes}-\eqref{momstokes} using a viscosity field
$\eta(u_0)=2\eta_0\|\DD(u_0)\|_F$ for a given $u_0$. This problem is
indeed a ``self-adjoint'' problem (the underlying operator is linear and symmetrical with respect
to $\vu$).\\

In general, the procedure  consists of calculating a mechanical equilibrium based
on the  complete non-linear system to obtain a converged $u_0$ and the
gradient is then obtained by simply  transposing the
final computed state. This method applied to the full-Stokes problem can
be  found in  \citet{morlighem2010}.\\

In  the automatic differentiation context, this approximation
is equivalent  to retain, in the reverse
accumulation process, only the gradient computed from the final evaluation of the function
$\Phi$. The quality of such an approximation is thus questionable and will strongly depend
on the problem one considers and the required accuracy on the gradient.\\

The quality  of this  approximation (compared  to the exact  adjoint state)  for parameter
identification is  assessed by  \citet{goldberg2011} for depth-integrated  shallow-ice type
equations but has never been treated for the full-Stokes equations.\\

\subsection{The continuous adjoint system }
\label{adjeq}
Before the numerical assessment of the ``self-adjoint'' approximation it seems relevant to
look into the continuous adjoint equation system  in order to highlight the terms that are
being ignored by the approximation and to estimate their weight in the complete adjoint system.\\

Omitting the lateral boundaries, the adjoint system of the full-Stokes
problem \eqref{debstokes}-\eqref{finstokes} is (see \textit{e.g.}
\cite{Petra2012}): 
\begin{eqnarray}\label{eq1}
-\di(\underline{\bs{\Sigma}}) &=&0 \mbox{ in } \Omega,\\
\di(\vv)&=&0 \mbox{ in } \Omega,\\
\underline{\bs{\Sigma}} \bs{n} &=& \vu_s^{obs}-\vu \mbox{ on } \Gamma_s,\\
\nonumber \Sigma_{nt} &=&\beta^{1/m}\left(|\vu_{\tau}|^{\frac{1-m}{m}}\vv_{\tau}+\right. ,\\
& & \left.(m-1)|\vu_{\tau}|^{\frac{1-3m}{m}}(\vu_{\tau}  \otimes \vu_{\tau})\vv_{\tau}\right) \mbox{ on }\Gamma_{fr}\label{fric},\\
\vv\cdot\bs{n}&=&0\mbox{ on }\Gamma_{fr},\label{eqn}
\end{eqnarray}
where $\vv$ denotes the adjoint velocity and $\underline{\bs{\Sigma}}$ the
adjoint stress tensor. The quantity $\Sigma_{nt}$ is defined in the
same way as $\sigma_{nt}$ (see equation \eqref{sigmant}). The adjoint stress
tensor is written:
\beq\label{adjstress}
\underline{\bs{\Sigma}}=2\eta(\vu)\left(\textrm{I}+\frac{2(1-n)}{n}
\frac{\DD(\vu)\otimes\DD(\vu)}{\|\DD(\vu)\|_F^2}\right)\DD(\vv)-\Id q
\eeq
with $q$ denoting the adjoint pressure, $\textrm{I}$ the fourth-order
identity tensor applied to order two tensors, $\Id$ the second order
identity tensor and $'\otimes'$ the tensor product.\\

By construction, this problem is a linear problem in $\vv$ and depends on the forward
velocity $\vu$. The method to derive the adjoint system associated to
any non-linear elliptic problem can be found in \eg \cite{coursjerome}.\\

First, the non-linearity of the forward problem appears in the
definition of the adjoint stress given in equation
\eqref{adjstress}. The norm of the term
$\frac{\DD(\vu)\otimes\DD(\vu)}{\|\DD(\vu)\|_F^2}$ is simply one
(since $\|\DD\otimes\DD\|=\|\DD\|_F\times\|\DD\|_F$ given a consistent
choice of the fourth-order tensor norm with the Frobenius matrix
norm) and the norm of the identity tensor is known to be greater or
equal to one (and typically equal to one for the \textit{sup}
norm). The linearity assumption of the "self-adjoint" method leads
to set $n=m=1$ in the adjoint system \eqref{eq1}-\eqref{eqn}. It then leads
to the dropping of a term that is comparable
to the one that is kept, for $\frac{1-n}{n}$ close to one ($2/
3$ for $n=3$). It logically follows that the greater the non-linearity
(the  value  of  $n$), the  greater  the  non-linear  contribution,  and the  coarser  the
``self-adjoint''approximation.\\ 

The  other  non-linearity   comes  from  the  non-linear  friction   law  and  appears  in
equation \eqref{fric}. A similar calculation leads to a similar conclusion: for $m>1$, the
norm  of  the terms  that  are  being dropped  by  the  ``self-adjoint'' approximation  is
comparable to the one being kept.\\

Let us point out that,
in equation \eqref{fric}, for larger values of $m$ (representing
hard-rock sliding or mimicking Coulomb friction), the nonlinear contribution is no longer
comparable to  the linear  part and  becomes dominant due  to the  factor $(m-1)$,  and to
neglect the nonlinear terms is most certainly unsuited.\\

These  observations are  clearly  retrieved  numerically in  the  gradient test  performed
hereafter  (see  Figure \ref{gradtestc5})  which  shows  a relative  error  around  $1$  for the  ``
self-adjoint'' approximation.

\subsection{Numerical evaluation of the ``self-adjoint'' approximation}
\label{praprecsadj}

We consider the  flow described in section  \ref{formod}.  The domain
is a  parallel sided slab on an inclined bed with an aspect ratio of $1/10$  on a $10\%$ slope. The
friction condition at the bottom is given  by \eqref{fric1} with a constant $\beta$ and an
exponent  $m=3$.  A  stationary free  surface flow,  uniform with  respect to  $x$,  is thus
obtained.

The cost function $j$ used here corresponds to the one defined by \eqref{costfunc}
withtout regularization: \beq\label{costogt} j=j(\beta;0) =
\frac{1}{2} \int_{\Gamma_s}
\left|u(\beta,\overline{z})-u_s^{obs}\right|^2~d\overline{\bs{x}} \eeq

where the observations $u^{obs}$ are the horizontal velocities at the surface $\Gamma_s$,
$(\overline{x},\overline{z})$ designates the mean-slope frame and the control variable
is the discrete friction coefficient field $\beta$.

\begin{figure*}[t] 
\begin{center}
\includegraphics[scale=0.32]{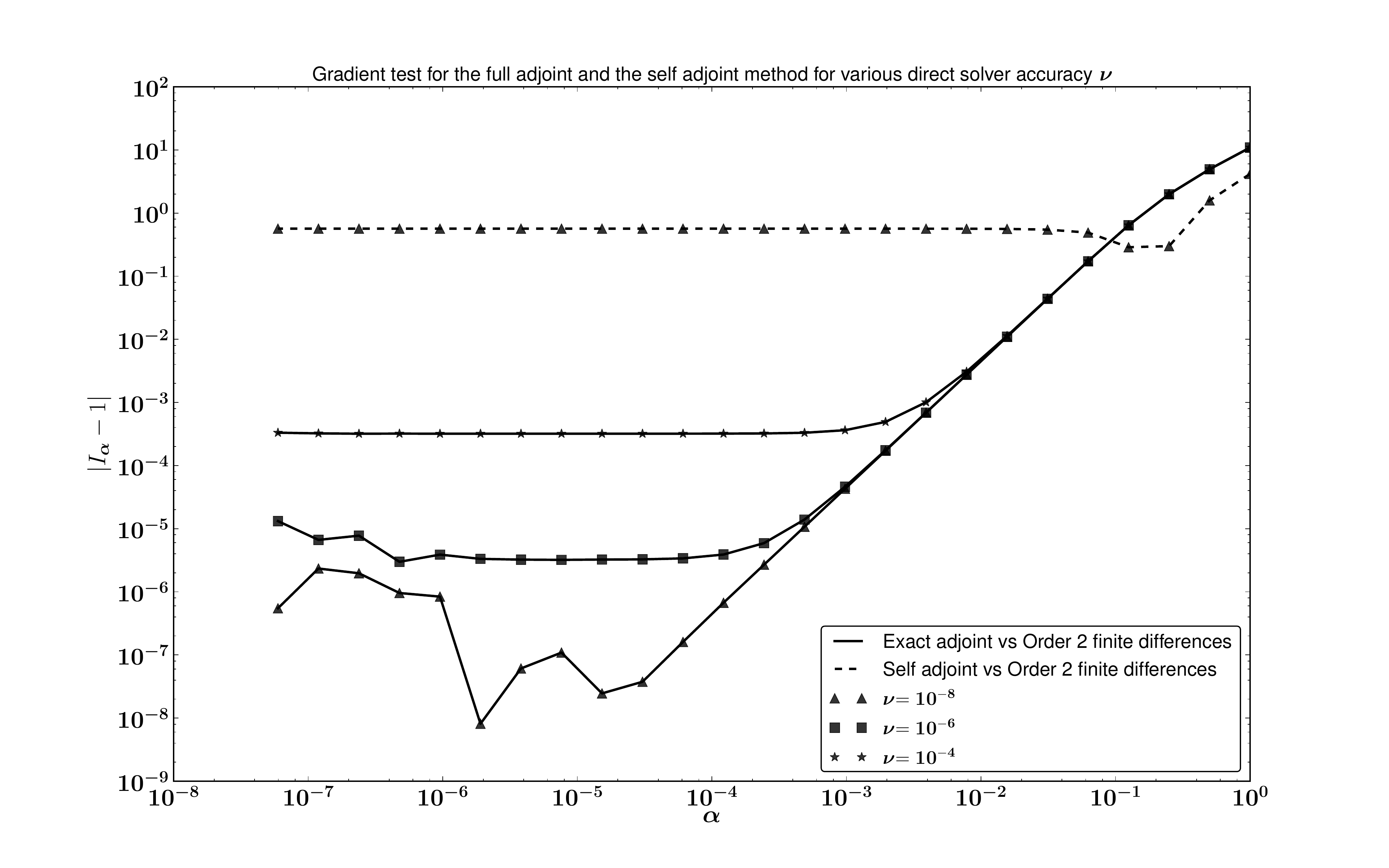}
\end{center}
\caption{Gradient  test for  the  full  adjoint method  the  ``self-adjoint'' method  for
  various levels of precision $\nu$ of  the forward solution. The quantity $I_{\alpha}$ is
  defined by \eqref{ialpha2}.}
\label{gradtestc5}
\end{figure*}

The gradient tests  carried out for the ``self-adjoint'' and  full adjoint methods, using
cost  function \eqref{costogt},  are plotted  in  Figure \ref{gradtestc5}.  The tests  are
performed    for    various    levels    of    precision   of    the    forward    problem
$\nu=\|u_{k+1}-u_k\|/\|u_k\|$ in order to quantify the best attainable precision by the
adjoint problem with  respect to $\nu$. This precision is explicitly  given to the direct
solver through a  convergence threshold for the nonlinear loop but can  be seen as the available  accuracy on the
data $u^{obs}$ ; a direct solution accuracy $\nu=10^{-4}$ mimics data presenting a noise of $0.01\%$. The
use of unnoisy data helps to preserve the theoretical constant rate
decreasing error of the gradient test, thus validating the method.\\

 The gradient test compares  the gradient computed  by the adjoint
code to a reference gradient. For these  tests, the reference gradient is obtained using a
centered finite  difference approximation (of order  $2$) computed for a  precision on the
function evaluation  of $10^{-12}$.  This precision being  considerably higher  than those
considered for the  solution of the forward problem, the  finite difference gradient plays
the role of an ``exact'' value (see section \ref{gradtestsec}).\\

The full adjoint method shows the expected theoretical behavior. We recover the slope of $2$ (in
logarithmic  scale)  associated with  the  order of  convergence  of  the finite  difference
approximation \eqref{central_FD}. Figure \ref{gradtestc5} thus shows that the precision of
the adjoint state is of the same order as the one of the direct solver.

On  the  contrary,  the  precision  of  the  gradient  provided  by  the  ``self-adjoint''
approximation is rather limited. The best reachable precision, as
expected from the continuous adjoint system analysis, is slightly smaller than $1$
irrespective of  the direct  solver precision  $\nu$ (and thus,  only one  gradient test
curve is plotted  in Figure \ref{gradtestc5}, for the case  $\nu=10^{-8}$, $\nu$ being the
precision of the forward solution).\\

The ``self-adjoint'' approximation used within  a parameter identification process is thus
not  able to compute  an accurate  gradient.  However, as  further discussed  thereafter,
numerical  tests  demonstrate  a  certain  ability for  this  approximation  to  partially
reconstruct the friction  coefficient (for a computational cost well below  the one of the
full adjoint method in the automatic differentiation context).
Nevertheless,  significant  weaknesses for  the  reconstruction of
high frequencies  as  well  as  the  reconstruction  of the  main  frequency  of  the  friction
coefficient  signal, specifically for  extreme situations  of sliding  (very slow  or very
fast), are brought to the forefront.

\subsection{Adjustable adjoint accuracy \& truncation of the reverse accumulation}
\label{troncapp}

\begin{figure*}
\centering
  \includegraphics[scale=0.32]{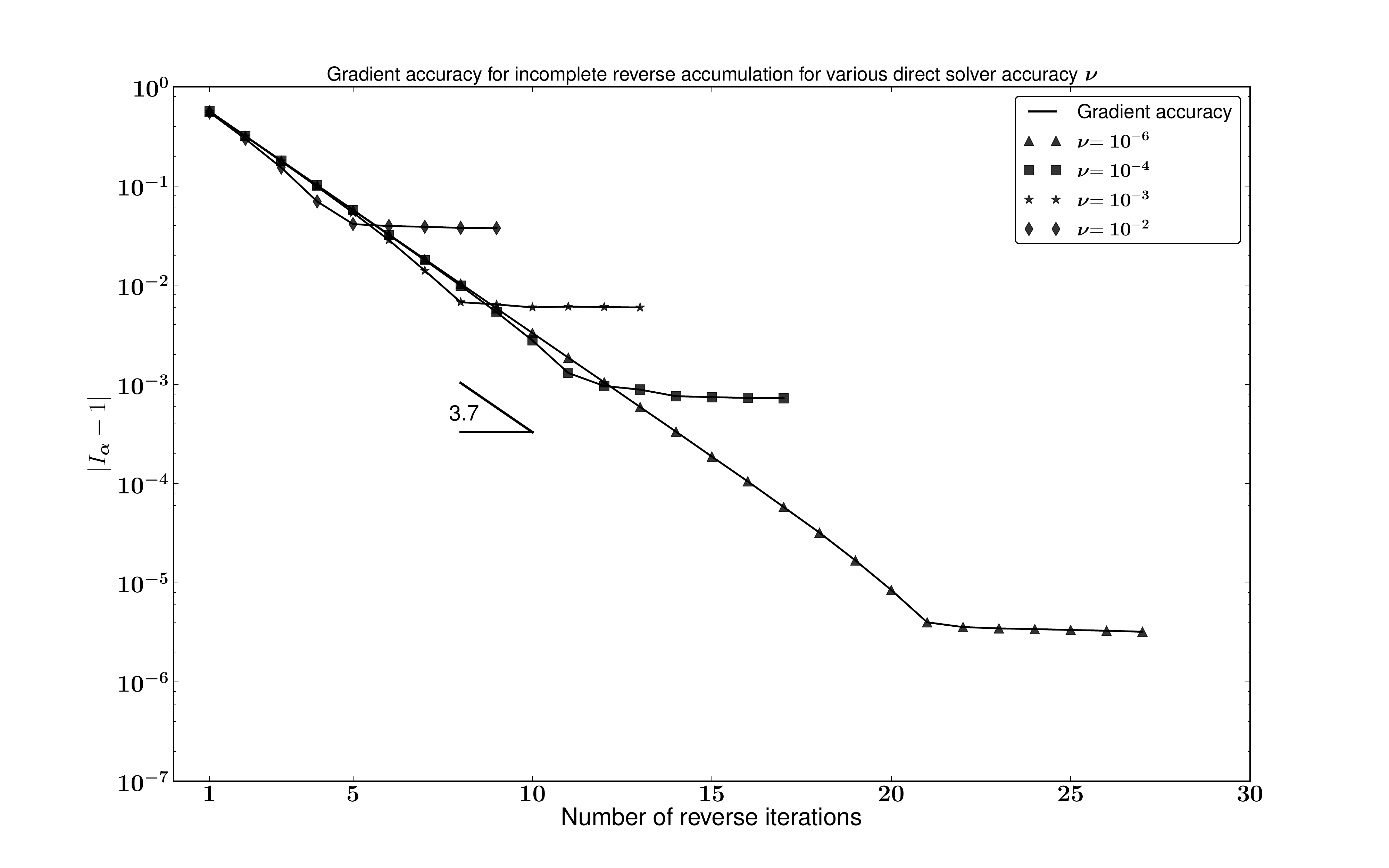}
\caption{Accuracy  of the  gradient  for incomplete  reverse  accumulation for  various levels  of
  precision of the direct solution $\nu$.}
\label{accrev}
\end{figure*}

This  section  focuses  on  the  effect  of  a  truncation  of  the  reverse  accumulation
process.  Figure  \ref{accrev}  plots  gradient  test results  obtained  for  a  truncated
evaluation of  the adjoint state.  To do so, the number of iterations of the adjoint loop is 
truncated from one to $N$, the total number of iterations performed by the direct solver. 
We thus obtain $N$ gradient tests providing every level of precision
for each  intermediary  adjoint  states between  the  exact  adjoint ($N$ iterations) and  the
``self-adjoint'' approximation. This  test is carried out for  various levels of precision
$\nu$ of  the direct solver. The  number of iterations $N$ performed by the direct  solver to
reach the required accuracy $\nu$ depends on this precision.\\

The results concerning the precision of the gradient presented previously are well
recovered (see Figure \ref{gradtestc5}). The  lowest precision, identical for every $\nu$
and equal to $0.6$, is  obtained from the ``self-adjoint'' approximation (corresponding to
$1$  reverse  iteration)  and  the  highest  precision is  reached  by
the  full  adjoint method (corresponding to the last point of each curve).\\

A linear  decrease of the error  (in logarithmic scale)  resulting in a slope of  $3.7$ is
observed. This behavior of the error is coherent with the  result of \citet{christianson1994} who
states that the computation of the adjoint state by reverse accumulation is equivalent to a fixed
point  computation.   In  the present  case,  we  have  a reverse  accumulation  algorithm
presenting a rate of convergence of $3.7$. Yet, the convergence speed of the forward fixed
point (not plotted  here) leads to a slope  of $3$.  The convergence of  the adjoint state
computation is therefore higher than the  one of the direct state computation. This result
explains  the  \textit{plateau} observed  for  the final  iterations;  indeed, a  faster
convergence of  the reverse accumulation algorithm  allows us to reach  the converged adjoint
state with fewer iterations.\\

Again, the accuracy of the ``self-adjoint'' approximation appears strongly limited and the
possibility of  an incomplete  method, intermediary between  the full
adjoint method and the
retention of only one iteration could bring an important gain of precision ;  taking into
account the linearly decreasing error (in logarithmic scale) leads to significantly
improved accuracy  for each  additional iteration retained  during the computation  of the
adjoint state.

Furthermore, the  faster convergence  of the reverse  accumulation algorithm compared  to the
direct solver allows, in  any case, to spare a few iterations during  the computation of the
adjoint  state without  any loss  of precision.  The number  of unnecessary  iterations is
likely to be strongly dependent on the situation and must be studied in every case.\\

For the  present test case,  we observe  that the $5$  last iterations during  the reverse
accumulation  are useless whatever  the level  of precision  of the  forward run  (see the
\textit{plateau} in Figure \ref{accrev}). These $5$ last iterations correspond to the $5$
first iterations  carried out  by the  direct solver. Avoiding  the accumulation  of these
iterations  for the adjoint  state evaluation  amounts to  starting the  reverse accumulation
from  a residual  on  the forward  run  of $0.1$  (\ie a  relative  variation between  two
successive iterates  of $0.1$).  This observation, although  dependent on  the considered
case, can  be seen as  an empirical method  to define a criteria  on the number  of direct
iterations that should be accumulated to obtain the best accuracy on the adjoint state. In
the present case, it amounts to initiating the memory storage of direct iterations once the
direct solver residual is lower than $0.1$.\\

In a more general point of view, the threshold imposed on the
direct solver to limit the accuracy of the computed state is a quite
numerical artifice and should not be seen as a way of saving time, regardless
of the data precision. A reliable approach for real numerical
simulations could be to perform an accurate direct simulation but a
truncated adjoint in adequation with the level of noise on the
data. This adjustment could be made based on one gradient test which
allows to quantify the rate of convergence of the reverse accumulation
loop.

\section{Friction coefficient identifiability}

This section focuses on the practical  limits of identifiability of the friction coefficient
by both the full adjoint and the ``self-adjoint'' method.\\

The main  goal is  to draw conclusions  on the  possibility of using  the ``self-adjoint''
method (which brings an  important time and memory saving) and then  on the quality of the
results  it  provides  in  the  perspective  of  realistic
identification  of  the  friction coefficient.  The  quality of  the  results  is evaluated  in  terms  of frequencies  and
amplitudes of the reconstructed friction coefficients compared to the target ones.\\

As presented before,  the precision of the ``self-adjoint''  gradient is bounded, whatever
the level of precision of the direct solver  $\nu$. This level of precision can be seen as
an \textit{a priori} accuracy on the  data considered in the cost function. In view of a 
thorouh analysis of the invertibility capacities of an adjoint-based inverse method, only synthetic
data  are used in  the present  work. A  Gaussian noise  of level  $\delta$ is  thus added
\textit{a posteriori}  to emulate real  data. The precision  of the exact  adjoint gradient
depending on $\nu$ (and equivalently on the  level of noise $\delta$ on the data), we seek
to observe which value of $\nu$ is required to observe the limit of precision of the
self-adjoint method.\\

To  this end,  we consider  three noise  levels $\delta$  of $0.01\%$,  $0.1\%$  and $1\%$.
representing referred as very low, low and realistic noise.  Although,
a realistic level of noise depends on many aspects, the use of GPS
techniques and InSAR velocity measurements can provide this type of
precision (\cite{king2004}, \cite{joughin2010}, \cite{rignot2011}).\\

In all cases,  the final cost reached by  both methods is not sufficient  enough to inform
their precision,  especially for noise level greater  or equal to 1\%  (which is typically
the case for real data). This means  that one cannot draw conclusions about the quality of
the ``self-adjoint'' approximation  solely based on a comparison of  the costs provided by
both methods.

On  the contrary  the frequency  analysis suggests  that an  identical final  cost  is not
equivalent to an  identical inferred friction coefficient. It  demonstrates that this type
of inverse problem is ill-posed, which can be seen as an equifinality issue (\textit{i.e.}
an identical  state, and consequently  an identical cost,  can be obtained  with different
sets of input parameters).

It is important to point out that the poorer the data (or similarly the greater the noise), the
stronger the equifinality.\\

In what follows, we first consider the idealized case of a
quasi-uniform flow on an inclined parallel sided slab with very low
and low noise level in order to highlight the numerical limits of the 
``self-adjoint'' method.

We then  perform pseudo-realistic, spatially  variable, flow experiments  with a realistic
noise for various density of the surface data. All the identifications presented hereafter
use, as an initial guess for the  friction coefficient, the average value $a$ of the target
coefficient.  The optimization procedure stops  when the  three following  criterions are
achieved: a relative  variation of the cost smaller than  $10^{-8}$, a relative variation
of the norm  of the gradient smaller than  $10^{-4}$ and a relative variation  of the norm
of the inferred friction coefficient smaller than $10^{-4}$.\\

\subsection{Quasi-uniform flow}

The following  experiments are  performed on the  same inclined parallel sided
slab as  in section
\ref{praprecsadj}.  A non-linear  friction law, defined by \eqref{fric1}  is considered at
the bottom  with an exponent $m=3$. The  target friction coefficient, variable  in $x$, is
given by:

\beq\label{fric4f}
\beta_r^N(x)=a+\frac{a}{2}\sin\left(\frac{2\pi
    x}{20dx}\right)+\frac{a}{5}\sum_{i=1}^N f_i(x) \eeq with

\beq
f_i(x)=\sin\left(\frac{2\pi x}{w_i dx}\right) \mbox{ with } w_1=10,~w_2=4,~w_3=2,
\eeq

 and by extension, we set: 

\beq
\bs{f_0}(x)=\sin\left(\frac{2\pi x}{w_0 dx}\right) \mbox{ with } w_0=20.
\eeq

The quantity  $a$   is  the   average  value  of   the  friction coefficient   in  $\mbox{Pa}\cdot
\mbox{s}\cdot\mbox{m}^{-1}$ and $dx=0.2$m denotes  the length of a basal  edge or, in other words,  the sharpness of
the bedrock discretization.\\

\begin{figure*}[t]
\centering
  \includegraphics[scale=0.26]{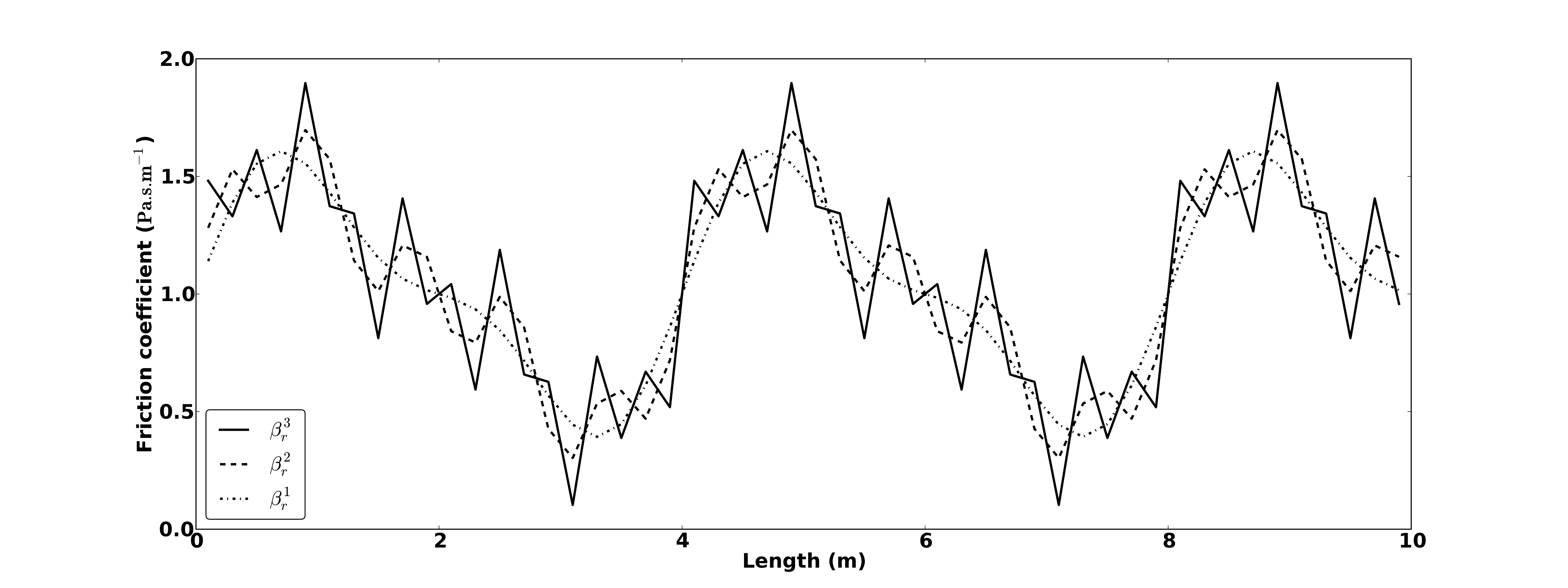}
\caption[Friction  coefficient $\beta$]{Friction  coefficient $\beta_r^n,1\leq  n  \leq 3$
  given by \eqref{fric4f} with $a=1$.}
\label{betari}
\end{figure*}

We  set  $\beta_r=\beta_r^3$, the  friction  coefficient resulting  from  the  sum of  $4$
frequencies corresponding to wavelengths of $20$,  $10$, $5$ and $2$ edge length $dx$. The
low frequency $f_0$ represents a carrier  wave for the $3$ higher
frequencies $f_i, i \in \llbracket 1,3 \rrbracket$. In  terms of
thickness of  the  domain $h$ (here constant and equal to $1$m, see
Table \ref{params1}),  frequencies $f_i,i \in \llbracket 1,3 \rrbracket$
correspond to wavelengths of $4h$, $2h$, $0.8h$ and $0.4h$ respectively. The coefficients 
$\beta_r^N,N \in \llbracket 1,3 \rrbracket$ are plotted in Figure \ref{betari} for  the case $a=1$. These
properties are summarized in Table \ref{params1}.

\begin{table*}
\begin{center}
\newcolumntype{R}{>{\raggedleft\arraybackslash}X}
\begin{tabularx}{0.64\textwidth}{X||c|c|c|c}
& $\bs{f_0}$ & $f_1$ & $f_2$ & $f_3$ \\ 
\hline\hline 
Wavelength w.r.t $h=1$m (thickness)& $4h$ & $2h$ & $0.8h$ & $0.4h$ \\ 
\hline
Wavelength w.r.t. $dx=0.2$m (edge length) & $20dx$ & $10dx$ & $4dx$ & $2dx$ \\
\hline
Wavenumber w.r.t. $L=10$m (domain length) & $2.5$m\textsuperscript{-1} & $5$m\textsuperscript{-1} & $12.5$m\textsuperscript{-1} & $25$m\textsuperscript{-1} \\
\end{tabularx}
\end{center}
\caption{Characteristics of signal $\beta$ given by \eqref{fric4f}.}
\label{params1}
\end{table*}

The flow is  uniform when the friction coefficient  is constant along the domain  and can be
described as quasi-uniform when the friction coefficient is given by \eqref{fric4f}. \\

We seek to determine  the level of spatial variability of the  friction coefficient the full
adjoint and the ``self-adjoint'' methods  can provide through the identification process, based
on surface velocity observations, with respect to the degree of slip.  The degree of slip
depends  on the  value  of parameter  $a$  and will  be described  thereafter  in terms  of
\textit{slip ratio}  $r$. The slip ratio  is a dimensionless quantity  that quantifies how
slippery  the bedrock  is. It  is calculated  as the  ratio of  the mean  sliding velocity
$\overline{u_b}$ to the difference  between mean surface velocity$\overline{u_s}$ and mean
basal  velocity   $\overline{u_b}$  (cf.   \citet{Hindmarsh2004}).  It  leads   to:  \beq
r=\overline{u_b}/|\overline{u_s}-\overline{u_b}|. \eeq

A slip  ratio $r=1$ represents a situation  where  half of the surface velocities are
 attributed to sliding and half are attributed to deformation.

We consider $6$ different slip ratios ranging from very high friction (close to adherence)
to very rapid sliding. The slip  ratios $r=0.005$, $r=0.05$ and $r=0.5$ can be described as
moderate sliding and the slip ratios $r=5$, $r=50$ and $r=500$ as rapid sliding.\\

\begin{figure*}[t]
\centering
  \includegraphics[scale=0.32]{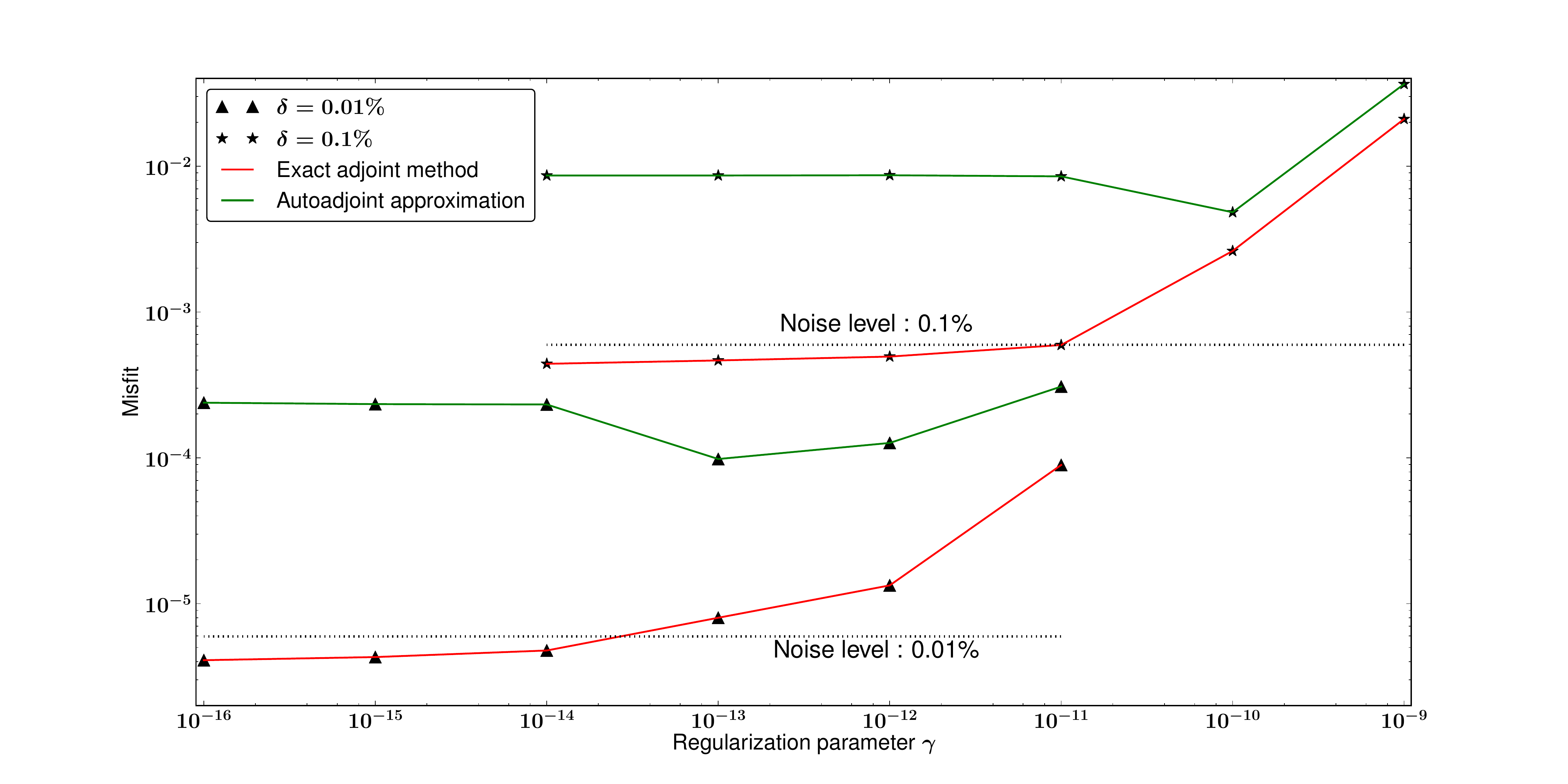}
  \caption{Application  of Morozov's  discrepancy principle.   Final discrepancy  (or
    misfit)  $j_1=\int_{\Gamma_s}\left|u_s(n)-u_s^{obs}(n_t)\right|^2\dx$ with  respect to
    the value  of the regularization  parameter $\gamma$.  The horizontal  line represents
    the level of  noise corresponding to the optimal discrepancy  obtained from the target
    coefficient.}
\label{morozovcanal}
\end{figure*}

In order to highlight the limitations of the ``self-adjoint'' approximation, the
 identifications of  $\beta$  performed hereafter consider noise levels  $\delta=0.1\%$ and
$\delta=0.01\%$   on   the  surface   velocity   data.  Let   us   point out  that   the
``self-adjoint'' method provides  very similar  results to the  full
adjoint one in  terms of final  cost  when  $\delta=1\%$  (not   plotted  in  Figure  \ref{morozovcanal})  and  the
distinction between both methods clearly appears for lower noises.\\

The cost function is defined by \eqref{costfunc}.
The tuning of  the regularization parameter $\gamma$ is achieved  according to the Morozov
discrepancy principle (see section \ref{morozovsec}). We plot in Figure \ref{morozovcanal}
the application  of this method to  the identifications performed with  both methods (full
adjoint and ``self-adjoint'') in the case of an intermediate friction ($r=0.5$). The
corresponding curves for other slip ratios are identical and consequently not plotted.\\

Figure  \ref{morozovcanal} clearly demonstrates the inability, for the
“self-adjoint” method, to provide a gradient  for
sufficiently low noise. For noise levels $\delta=0.1\%$ and
$\delta=0.01\%$, the ``self-adjoint'' gradient does not allow the
optimal misfit to be reached. Therefore, in these situations, the 
``self-adjoint'' approximation is theoretically not valid. However,
as we will see, the ``self-adjoint'' method shows a certain
ability to retrieve the target parameter. This  observation  is
independent of the degree of slip.\\

In order to study the effects of the approximation on the gradient computation, we compare, in
the following,  the friction coefficient inferred  by both methods  for $\delta=0.01\%$ and
$\delta=0.1\%$.\\ 

\begin{figure*}[h!]
 \includegraphics[scale=0.9]{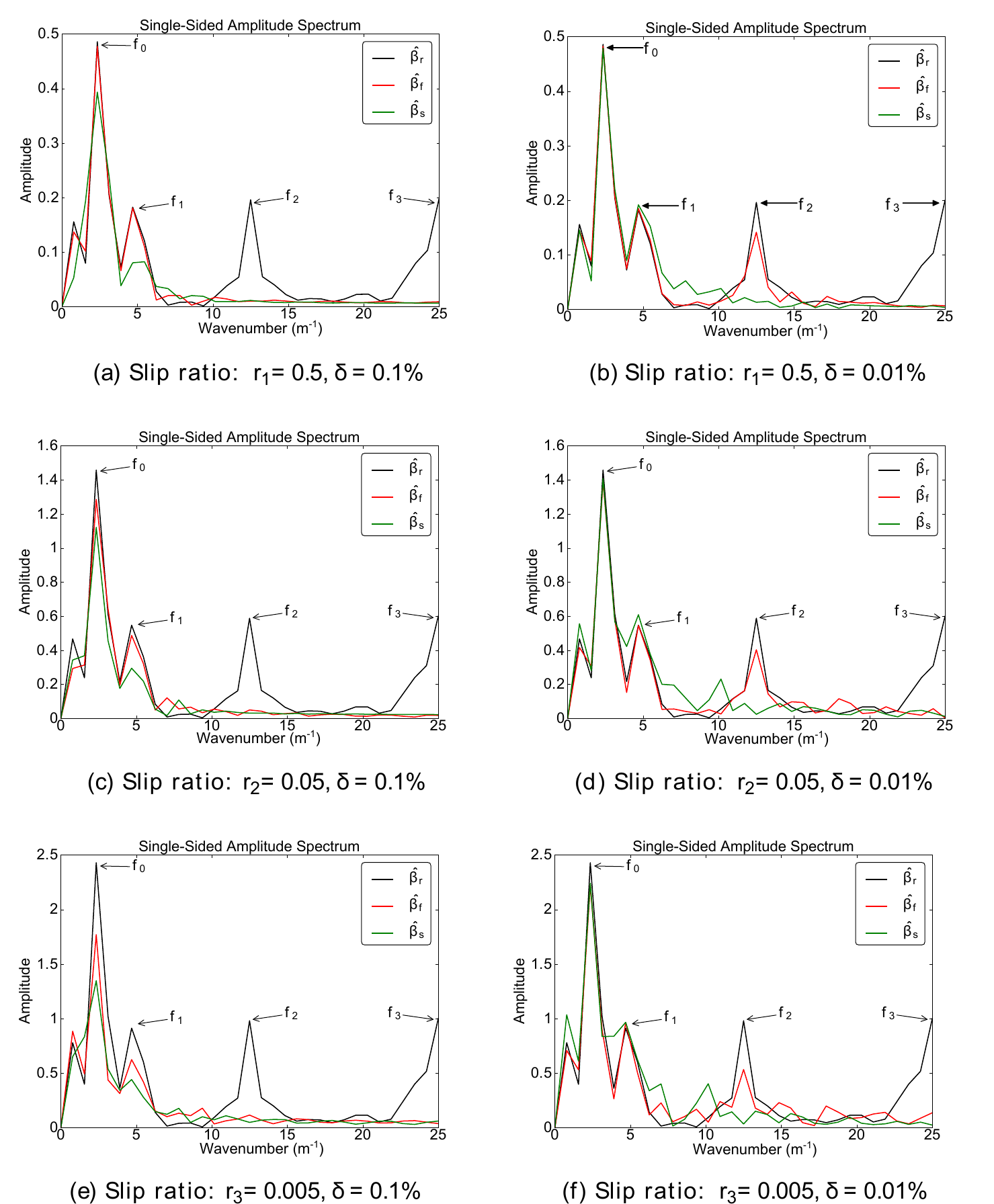}
 \caption  {Discrete Fourier  Transform  of inferred  friction  coefficient $\beta_f$  and
   $\beta_s$ and of the target friction coefficient $\beta_r$. Moderate sliding.}
 \label{fftcanal}
 \end{figure*}

\begin{figure*}[h!]
\includegraphics[scale=0.9]{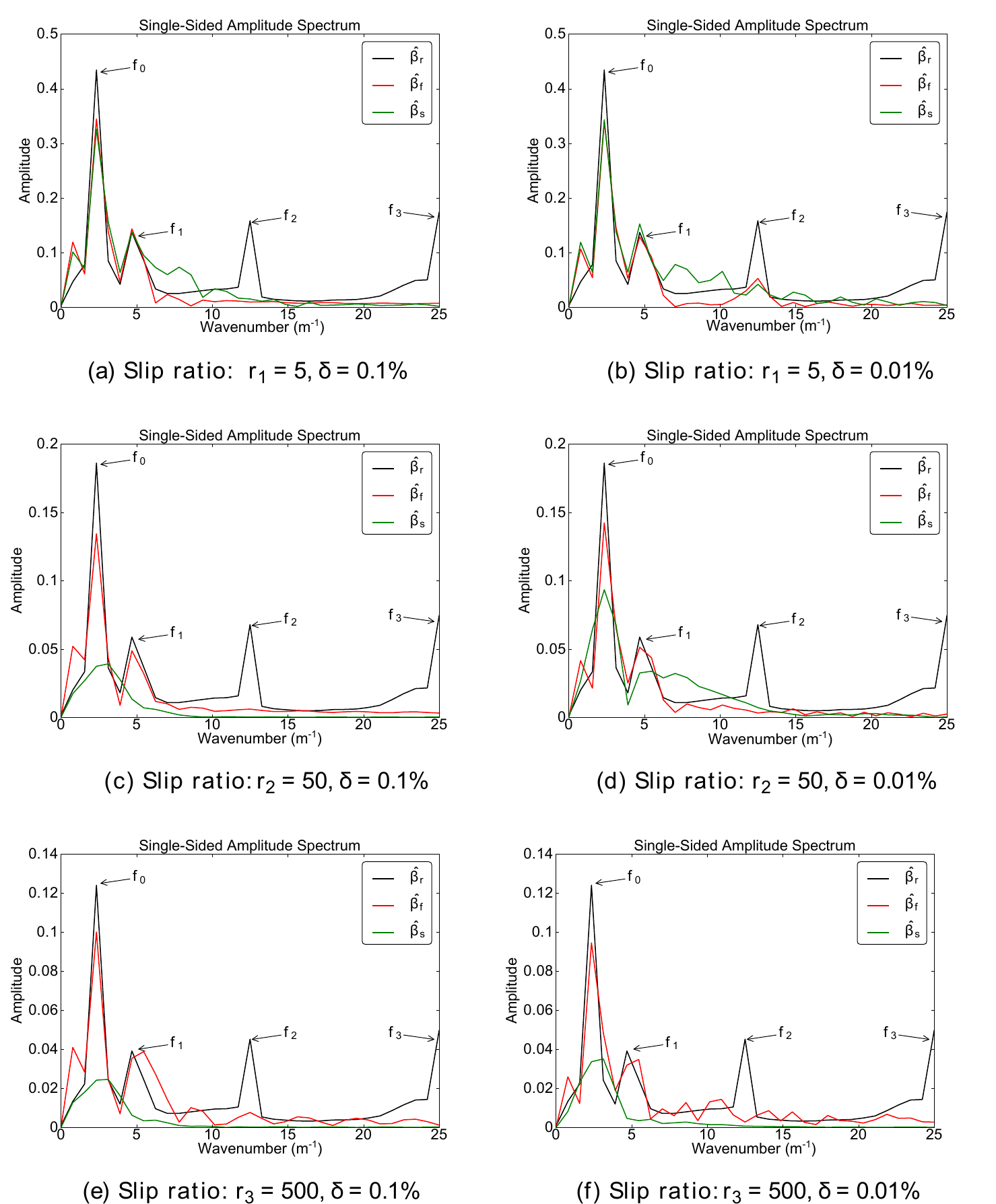}
 \caption {Discrete Fourier  Transform  of inferred  friction  coefficient $\beta_f$  and
   $\beta_s$ and of the target friction coefficient $\beta_r$. Rapid sliding.}
 \label{fftcanal2}
 \end{figure*}

The best inferred friction coefficient (according to Morozov) are noted $\beta_f$ for the 
\textit{full-adjoint}  and   $\beta_s$  for  the   \textit{self-adjoint}.  The  quantities
$\widehat{\beta_f}$ and  $\widehat{\beta_s}$  thus   denote  their  associated  Discrete  Fourier
Transform (DFT).  We denote by $\widehat{\beta_r}$  the DFT associated to  the target friction
coefficient \eqref{fric4f}.  The single-sided amplitude spectrum of the DFTs $\widehat{\beta_r}$, $\widehat{\beta_f}$ and $\widehat{\beta_s}$
obtained  for  the   three  small  slip  ratios  (moderate  sliding)   are  plotted  in  Figure
\ref{fftcanal} and those obtained for the three high slip ratios (rapid sliding) are
plotted  in  Figure \ref{fftcanal2}.  The  amplitude spectrum  plots  the  modulus of  the
complex Fourier coefficient multiply by two, providing the original amplitude of
the  frequencies   of  the  signal  (approximated   by  the  sharpness   of  the  sampling
frequency). The abscissae have been rescaled according to the discretization of the bedrock $dx$
and the length of  the domain $L$ in order to directly  provide the original wavenumber of
the frequencies.  All the signals  have been centered  (to have a  zero mean) in  order to
remove  the peak  corresponding to  the average.  Since the  zero mean amplitude  spectrum is
symmetrical, the  single-sided spectrum is plotted everywhere.  The single-sided amplitude
spectrum plotted in Figures \ref{fftmertz}, and \ref{fftmertzmask} are identically defined.\\

\paragraph{Moderate sliding}

One observes first that frequencies $f_0$  and $f_1$ (see Table \ref{params1}) are globally
well reproduced  by both methods for  $\delta=0.01\%$ whatever the slip  ratio. Namely the
carrier frequency $f_0$ is very well reconstructed by both methods and this property seems
desirable. The full  adjoint method shows a greater robustness  when identifying these two
low frequencies with respect to the  slip ratio whereas a noticeable deterioration for the
identification of frequency  $f_1$ occurs for the ``self-adjoint''  method when slip ratio
decreases. 

However, frequency  $f_2$ appears correctly captured  by the full adjoint  method while it
does  not appear  in the  spectrum of  the ``self-adjoint''  one. An  increased difficulty
in capturing this frequency occurs with slip ratio increasing.

Finally, the highest frequency $f_3$ does not appear in any of the spectrums of both $\beta_f$
and $\beta_s$ whatever the degree of slip.\\

For a noise level $\delta=0.1\%$, one  loses the ability to retrieve frequency $f_2$ using
the full adjoint  method. The identification  of frequency $f_1$  is accurately obtained  for the
slip  ratio $r_1=0.5$  but  we  observe a  deterioration  of the  result  when slip  ratio
decreases. The ``self-adjoint'' method captures almost none of frequency $f_1$ whatever the
slip ratio.

Concerning  the carrier  frequency,  one observes  difficulties  for the  ``self-adjoint''
method  to reconstruct  it accurately  even for  the slip  ratio $r_1=0.5$.  The frequency
distinctly appears on  the spectrum but only $80\%$ of the  target amplitude is recovered.
The decreasing  of the slip  ratio deteriorates, for  both methods, the  identification of
$f_0$. In the case $r_3=0.005$, the full adjoint method recovers $70\%$
of the target amplitude where the ``self-adjoint'' method recovers $50\%$.\\

\paragraph{Rapid sliding}
Again, low frequencies $f_0$ and $f_1$ are well retrieved with the full adjoint method for
every  noise level.  The carrier  wave reconstruction  is nevertheless  diminished (around
$80\%$ of the  target amplitude) compared to the moderate  sliding situation $r_1=0.5$ but
is stable with the  increasing of $r$. Similarly, frequency $f_1$  is rather well represented
by the full adjoint method for all the situations despite a  certain degradation with increasing
$r$. However, frequency  $f_2$ does not appear in any spectrum  irrespective of both slip
ratio and method, contrarily to the moderate sliding situations. Again, frequency $f_3$ is
never captured.  A small but noticeable  noise appears for  the case $r=500$ for  the full
adjoint method, particularly when $\delta=0.01\%$.\\

The ``self-adjoint'' method shows a relatively  good reconstruction of $f_0$ and $f_1$ for
the  case $r=5$ but  introduces noise between  frequencies $f_1$  and $f_2$.  A strong
deterioration  of  the reconstruction  occurs  when  $r$ increases  ;  for  a noise  level
$\delta=0.1\%$, the ``self-adjoint'' identification is almost unable to recover the signal
for $r\geq 50$.

\paragraph{Assessments} From these observations, we draw the following conclusions. Firstly, the
degree of slip of the target plays a  strong role for the limit of identifiability of the
friction coefficient  in  terms of  frequencies  ;  a  smaller  slip  ratio induces  a  lower
sensitivity of the  flow to the friction coefficient and consequently  a higher filtering on
the transmission of information from the bedrock to the surface.  \\

A  strong  friction  induces a  vertical  velocity  profile that is rather convex  with  velocity
gradients (shearing) mostly concentrated close to the bottom leading to a weaker transmission
of the information from the bottom to the surface. A similar
observation can be made from the sensitivity of the model to the
rheological constant $\eta_0$: the high sensitivity areas are
strongly correlated to the areas of high shearing.\\

Similar  to   strong  frictions,   low  frictions  also   reduce  the  quality   of  the
reconstruction.  This again  comes from  a reduced  sensitivity of  the flow  to  the friction
coefficient when  rapid sliding occurs however  this lower sensitivity  appears for different
reasons. Intuitively, the  case of a very low friction leads  to lower local topographical
effects and the resistance to the ice flow acts through an equivalent global topography at
larger scale.   This characteristic appears in  the explicit solution of  the uniform flow
\eqref{permsol}:  in order for  the mathematical  expression to  make sense  when $\beta$
tends to $0$, it requires the slope parameter $\theta$ to tend to $0$ as well. This
phenomena is  physically observed:  in the  presence of an  extended sub-glacial  lake, one
observes a signature  of this lake at the  surface as a very flat  surface topography over
the  lake. This interpretation  is retrieved  in the  normalized sensitivities  plotted in
Figure \ref{sens_mertz}.

These  two observations  support the  existence of  a numerical  identifiability
maximum for  the friction coefficient using  adjoint-based method ; the  best situation to
carry  out identifications corresponds  to the  intermediate friction  range where
sliding  effects  and  deformation  effects   on  the  dynamics  are  balanced  (typically
$0.5<r<5$).  The low  accuracy of  the ``self-adjoint''  gradient appears  to be  a strong
limitation in the case of rapid sliding ($r>5$).\\

For  the  current quasi-uniform  flow,  for  a noise  level  $\delta=0.1\%$,  a limit  on
identifiable  wavelength using the full  adjoint method, for  any degree  of slip,  is $2h$, where $h$ is the
thickness of the  domain. More accurate data could allow us to  infer higher frequencies in
the case of moderate sliding ($r\geq 0.5$).

For the ``self-adjoint'' method, for a slip  ratio $r\leq 5$, a wavelength of $4h$ is well
inferred and  a wavelength of  $2h$ is captured  for $r=0.5$ and  $r=5$. For a  slip ratio
$r>5$, the frequencies considered in the experiment are inappropriate.\\

In other  respects, a  tendency for the  ``self-adjoint'' method to  introduce non-physical
interferences  within the  inferred  coefficient for  very  low noise  appears.  This  non
desirable  phenomena  increases when  the  slip ratio  takes on extreme  values. Beyond  the
approximation aspect, one  can deduce a lack of robustness  of the ``self-adjoint'' method
for very low  noises.  It seems coherent with regards to the  low precision the ``self-adjoint''
gradient  provides. On  the contrary,  the full  adjoint method  provides a  less accurate
identification when  the slip ratio  goes away from  $1$ without introducing  non physical
effects in the
inferred parameter.\\

It  is of  interest to  notice that  the inability  to recover  frequency $f_3$  is  not a
numerical limitation  but a  limitation due to  the noise  on the data.   For sufficiently
accurate data, it is also identifiable using the second order exact adjoint method. \\

Similar experiments are performed in the next section for a pseudo-realistic flow ran on a
radar vertical profile of the grounded part of the Mertz glacier in Antarctica for surface
velocity data with different density and a $1\%$ noise.

\subsection{Real topography flow: the Mertz glacier}

\begin{figure*}[ht!]
\begin{center}
    \includegraphics[width=1.0\textwidth]{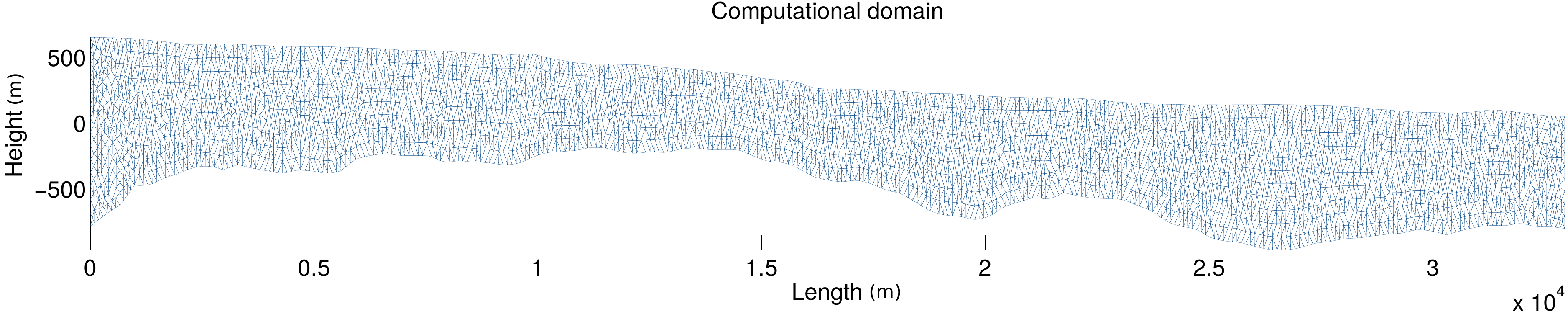}
    \caption{Vertical cut of the outlet glacier Mertz, Antarctica
      (topography profile from ICECAP 2010 within Ice bridge, provided
      by B. Legr\'esy, LEGOS, France) , x-scale $=2/5$}\label{mertzo}
\end{center}

\end{figure*}

The   flow  considered   in  this   section  is   identical  to   the  one   presented  in
subsection \ref{formod}. The computational domain is built from real field data; topography of the bedrock and of
the surface are bidimensional radar-sensed layers of the \textit{Mertz ice tongue} in East
Antarctica. These layers have been measured along a flowline of this outlet glacier
(American program ICECAP 2010, see \citet{greenbaum2010}). Our study focuses on the
grounded part of the glacier. The computational domain is plotted in
Figure \ref{mertzo}.\\

Synthetic data are obtained using the following friction coefficient:
\beq\label{fric4f2}
\beta_r^N(x)=a+\frac{a}{2}\sin\left(\frac{2\pi
    x}{50dx}\right)+\frac{a}{5}\sum_{i=1}^N f_i(x) \eeq with

\beq
f_i(x)=\sin\left(\frac{2\pi x}{w_i dx}\right) \mbox{ with } w_1=20,~w_2=10,~w_3=5,
\eeq

 and by extension, we set: 

\beq
\bs{f_0}(x)=\sin\left(\frac{2\pi x}{w_0 dx}\right) \mbox{ with } w_0=50.
\eeq

The quantity $a$ is the average friction coefficient and $dx=100$m is the bedrock edge length.
The  context of  a  non-uniform flow  on  a complex  topography allows  to  carry on  the
comparison between both  methods in the case  of a realistic flow simulation.  We can then
draw   practical   conclusions   on   the   validity   of   using   the   ``self-adjoint''
approximation. Frequency $f_0$  is a carrier wave with  $50dx$ wavelength corresponding to
$5h$, where $h\sim 1$km is the average thickness  of the domain. Frequencies $f_1$, $f_2$ and $f_3$
correspond then to wavelengths of $2h$,  $h$ and $h/2$ respectively, providing a situation
similar to the inclined slab test case (see Table \ref{params2}).\\ 

\begin{table*}
\begin{center}
\begin{tabularx}{0.69\textwidth}{X||c|c|c|c} & $\bs{f_0}$ &
$f_1$ & $f_2$ & $f_3$ \\ 
\hline\hline 
Wavelength w.r.t $h=1$km (thickness) & $5h$ & $2h$ & $h$ & $0.5h$ \\ 
\hline
Wavelength w.r.t $dx=100$m (edge length) & $50dx$ & $20dx$ & $10dx$ & $5dx$ \\
\hline
Wavenumber w.r.t $L=33.3$km (domain length) & $6.6$m\textsuperscript{-1} & $16.6$m\textsuperscript{-1} & $33.3$m\textsuperscript{-1} & $66.6$m\textsuperscript{-1}
\end{tabularx}
\end{center}
\caption{Characteristics of signal $\beta$ given by \eqref{fric4f2}.}\label{params2}
\end{table*}

\begin{figure*}[h!]
\begin{center}
    \includegraphics[width=0.8\textwidth]{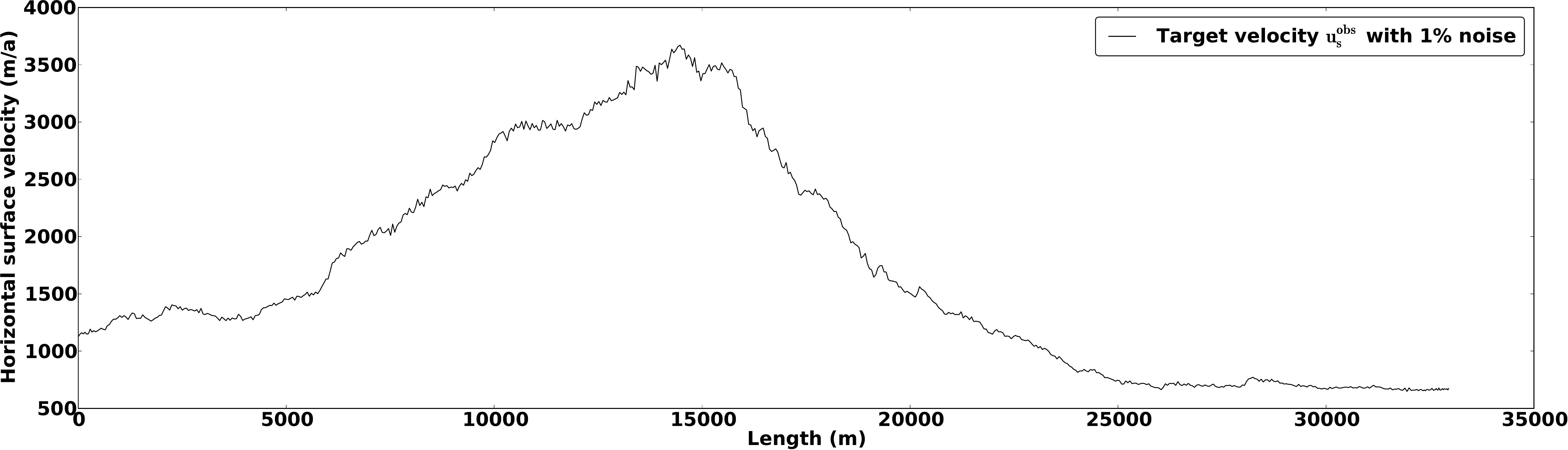}
    \caption{Horizontal surface velocities used as synthetic data in
      the case $r=5$ perturbed with a 1\% noise}\label{usfig}
\end{center}

\end{figure*}

In the present case  of a non uniform flow with complex topography,  it is not feasible to
simulate an  average slip ratio $r=500$.  Given the important spatial  variability, we are
able to achieve  a maximum average slip ratio $r=50$. In  the following identification, we
consider only $5$ slip ratios ranging from $r=0.005$ to $r=50$. The
synthetic horizontal surface velocity perturbed with a $1\%$ noise are
plotted in Figure\ref{usfig} for the case $r=5$.

\begin{figure*}[t]\centering
\includegraphics[scale=0.34]{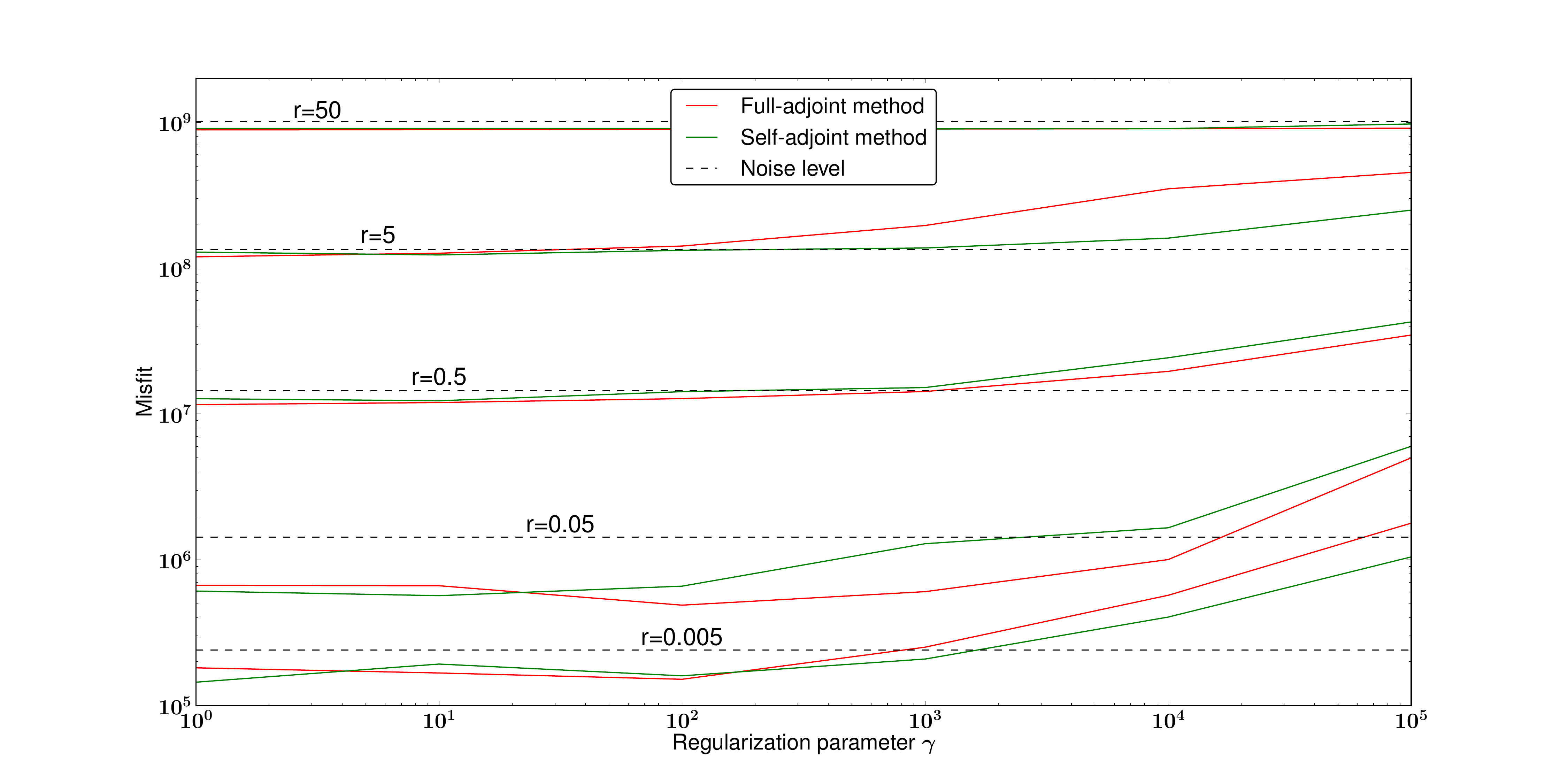}
\caption{Morozov's  discrepancy principle applied to slip  ratios $r=0.005$, $r=0.05$,
  $r=0.5$,  $r=5$ and  $r=50$ on  the realistic  flow. Absolute  values of  the discrepancy
  correspond to  the real value obtained  during the simulations. The  range of parameter
  $\gamma$ has been modified to remain between $1$ and $10^5$ for the sake of readability}
\label{morozovmertz}
\end{figure*}

The Morozov's discrepancy  principle applied to these $5$ situations  is plotted in Figure
\ref{morozovmertz}. 

The observed  behavior is similar to  the one previously  noted (but not plotted)  for the
idealized situation.   Both methods behave identically  in terms of cost  decreasing for a
$1\%$ noise  level on  the data.  In all cases,  they demonstrate  a robust  behavior that
provides  an  optimal  discrepancy  (according  to  Morozov).  The  expected  behavior  of
over-fitting (\ie to  reach a final misfit  smaller than the one computed  from the target
friction  coefficient with perturbed  data) for  $\gamma$ small  enough suggests  that the
gradient provided by both methods is \textit{a priori} accurate enough with regards to the
noise level (unlike the slab case with smaller noise, see Figure \ref{morozovcanal}).  The
peculiar behavior for the case $r=50$ where the discrepancy remains lower than the optimal
one regardless of the regularization parameter $\gamma$ is detailed hereafter.

\begin{figure*}[t]
 \centering
\includegraphics[scale=0.9]{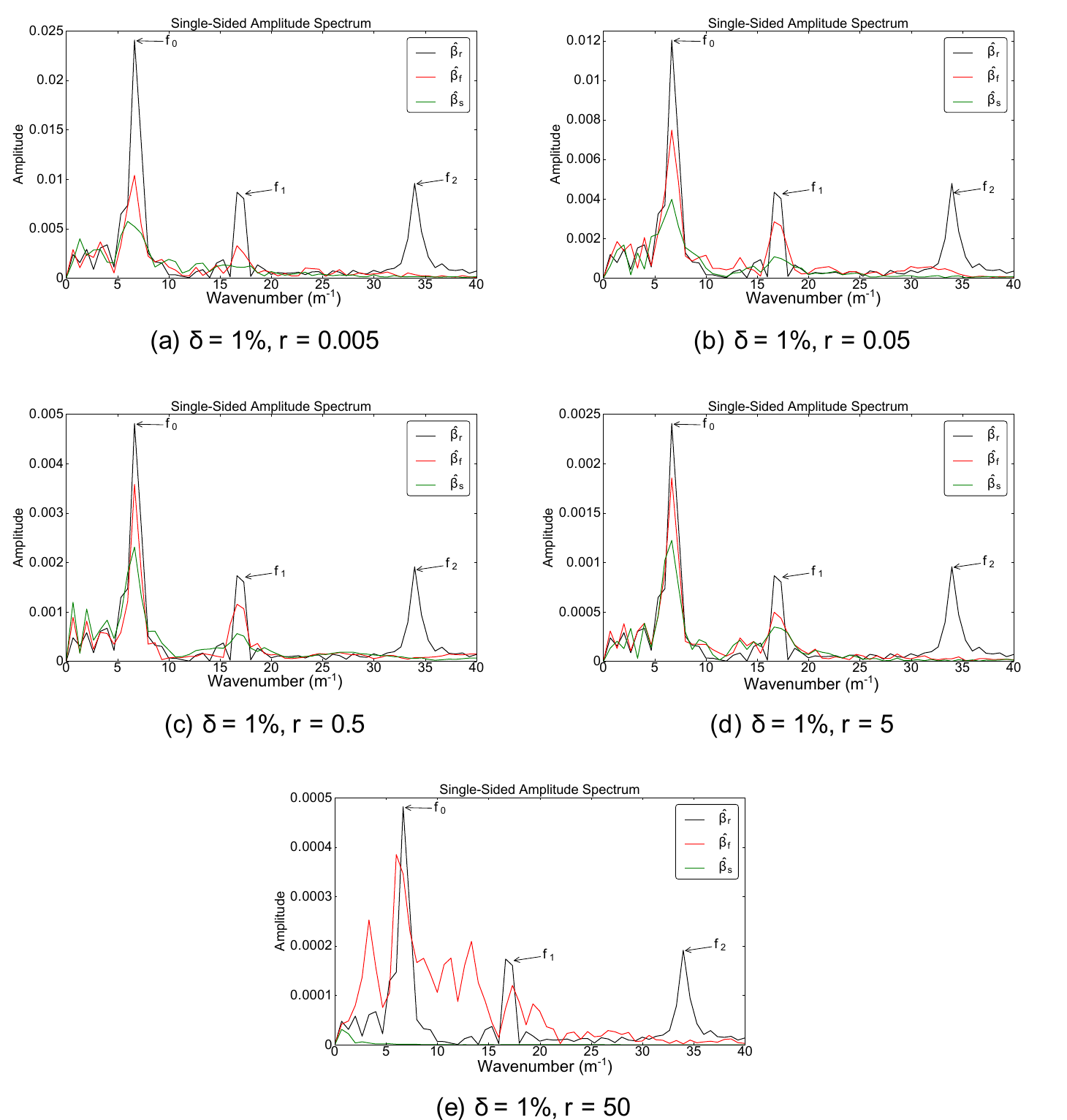}
 \caption
 {Discrete Fourier  Transform for inferred  friction coefficients $\beta_f$  and $\beta_s$
   and for the target  one $\beta_r$. Frequency $f_3$ is never captured  by any method and
   is  thus  not plotted  on  the  curves.  A noise  level  $\delta=1\%$  is used  in  all
   situations.} 
 \label{fftmertz}
 \end{figure*}

Figure \ref{fftmertz} plots the DFT of  the friction coefficients inferred by both methods
and of the  target coefficient \eqref{fric4f} for a  noise level of $1\%$ on  the data and
for the $5$ slip ratios $r$.

While the  wavelengths considered in  the friction coefficient \eqref{fric4f}  are similar
(in terms of thickness ratio) to those considered for the quasi-uniform test case, the use
of a higher noise on a non-uniform flow deteriorated the reconstruction at all levels.
The  carrier frequency  amplitude (of  wavelength $5h$)  is never  fully recovered  by any
method but clearly appears for $r \leq 5$. Likewise, frequency $f_1$ (of wavelength $2h$),
well  captured  in  previous simulations  by  the  full  adjoint  method, is  fairly  well
reconstructed only  for $0.05\leq r\leq 5$.  Again the ``self-adjoint'' method  is able to
recover it only partially. However, the interferences introduced by the ``self-adjoint''
method within  the inferred friction coefficient do  not appear anymore for  this level of
noise on the surface data.  It therefore seems coherent with the limited accuracy of the
gradient provided by this method.

As  a consequence,  the  chosen  frequencies for  these  simulations are  too  high to  be
recovered in this non-uniform flow with realistic data. Numerical
experiments using higher wavelengths in the friction coefficient show that an
accurate  reconstruction for any slip ratio can be  obtained, for  the full adjoint method, for  a carrier  wave of
wavelength  $10h$ and  a  perturbation of  wavelength  $5h$ ;  shorter  wavelengths are  not
accessible.\\

What is of further interest is that the full adjoint method brings, in all cases, an enhanced and
more faithful reconstruction of the friction coefficient for both the carrier wave and the
first perturbation.\\

The pattern of behavior of the rapid sliding case ($r=50$) is different compared to the other
cases.  The full  adjoint method  retrieves roughly  the carrier  frequency with  very high
interferences  (including   one  low  frequency  high  amplitude   interference)  and  the
``self-adjoint'' method does not capture any  information of the target signal in addition
to the initial guess.

\begin{figure*}[t]
  \includegraphics[scale=0.32]{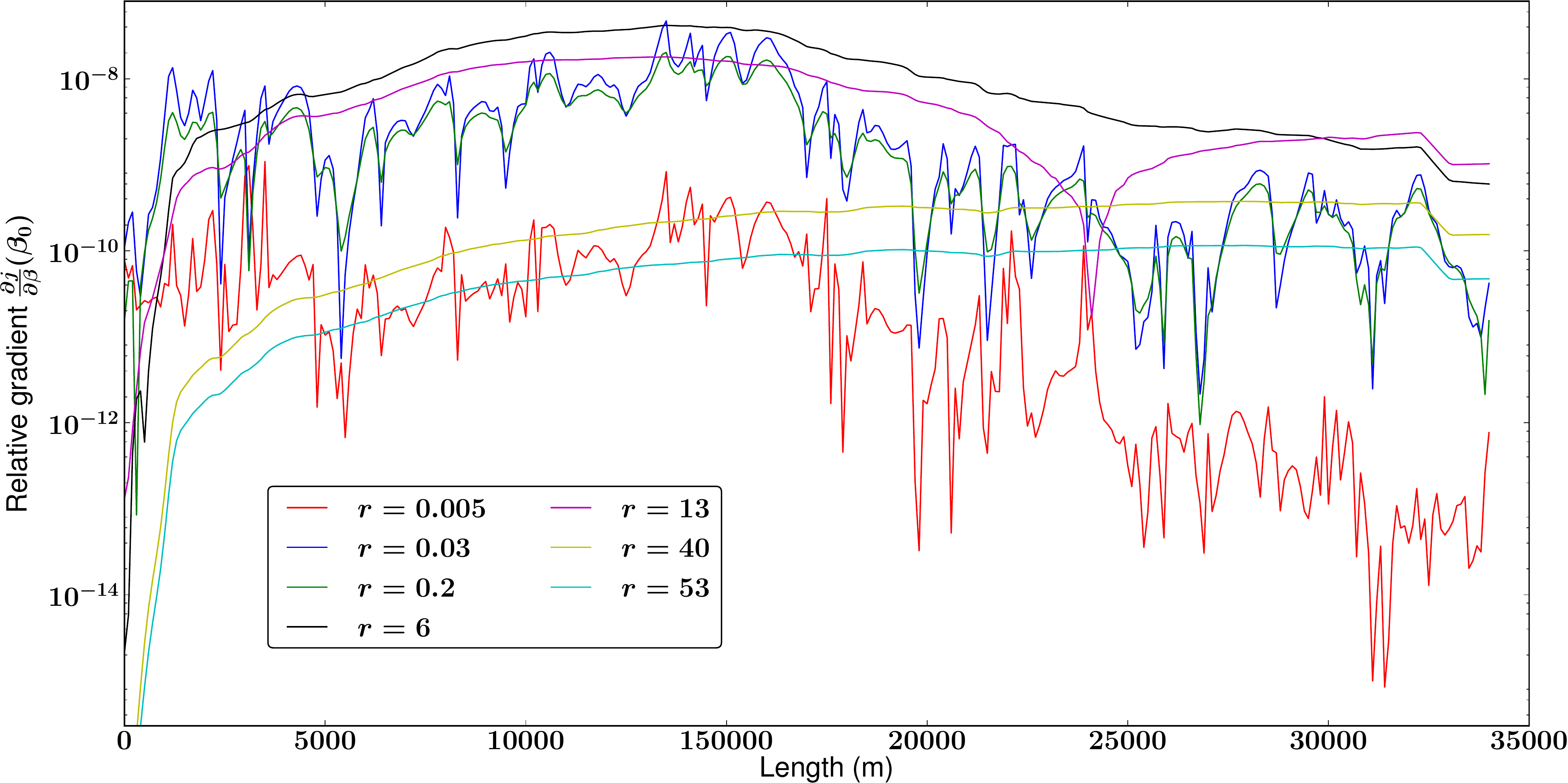}
  \caption{Relative sensitivities to the friction coefficient with respect to the abscissa
    $x$ for various slip ratios, evaluated around the average value $a$.}
\label{sens_mertz}
\end{figure*}

In order to understand this phenomena, we plot in Figure \ref{sens_mertz} the gradients
$\partial j/\partial \beta (\beta_0)$ with $\beta$ defined by \eqref{fric4f2} for several 
average value  $a$ of the friction coefficient, described in  terms of slip ratio  $r$. The
computed gradients are evaluated around $\beta_0=a$.

\label{adschap5}

Increasing the slip  ratio has a very clear effect on  the sensitivities. For slip ratios
$r<1$, the  sensitivities include the local  effects of the high  frequencies contained in
$\beta$, thus  providing a highly variable  gradient around an average  behavior. The fact
that the sensitivity decreases with $r$,  due to poorer information transmission between
the  bottom and  the surface,  is recovered.  It  follows that,  in the  cases $r<1$,  the
limitations in  the identification of all  the frequencies of the  friction coefficient come
from the precision on the data. 

The situations $r>1$  bring significantly smoother gradients. The  cases $r=6$ and $r=13$,
that still represent  moderate slip ratios, contain a certain  local variability but their
rather  smooth  appearance shows  a  strong correlation  with  the  global topography  (or
similarly the surface velocities, see  Figure \ref{mertzo} and Figure \ref{usfig}) and the
high frequencies of  $\beta$ seem already erased from the  gradient.  In these situations,
the main component  resisting the flow is more the large  scale (or equivalent) topography
than the friction itself.

For higher slip ratios, the topographical effects seem to vanish as well, and the gradient
only grows  from the inflow  boundary to  the outflow boundary  to reach a  maximum value
close to the right border. In the present case, one can deduce that the only effect
resisting the  flow is the cryostatic pressure  considered on the right  boundary.

A global decreasing  of the sensitivity with increasing $r$  is also observed, reinforcing
the existence of a  sensitivity peak for in-between $r$. For $r>1$,  it is not the quality
of the  data that  prevents an  accurate reconstruction  of $\beta$ but  the non-local
behavior of  the flow. When basal friction vanishes, it  does not embody more  than a
small fraction of the global resistance to the flow.   An extreme example is the progress of
an ice-shelf  on water where the  friction resistance is close  to zero. In the  case of a
tridimensional solution, stresses would be taken  over by lateral shearing.  In our case,
these effects  do not  exist and it  is the  hydrostatic pressure boundary  condition that
resists the  flow. These  clearly non-local effects suggest than the flow can only be globally controlled, thus limiting the
range of identifiable frequencies, regardless of data accuracy. Let us recall that, in
terms of absolute errors, a higher slip ratio leads to a smaller
absolute value of the friction coefficient and thus to a smaller
amplitude of errors. We also point out that the vanishing of the
sensitivities close to the left boundary is due to the Dirichlet
boundary condition.\\

These phenomena imply a strong equifinality for friction coefficient lower than a
certain   value.  This   observation  appears   in   the  Morozov's   curves  (see   Figure
\ref{morozovmertz}) for the  case $r=50$.  Indeed, the discrepancies  for both methods are
smaller than  the theoretical optimal one,  even for very  strong regularization ($\gamma$
large)  providing almost  constant $\beta$  around  $\beta_0$.  The  initial cost  itself,
evaluated for  a constant $\beta$ equal  to the average  value, is barely higher  than the
theoretical  optimal  cost. The  associated  minimization  problem  is ill-posed  and  the
Tikhonov  regularization on  the  gradient of  $\beta$  does not  allow  to overcome  this
problem.

For  the case  $r=50$ and  a  regularization small  enough (considering  that the  Morozov's
principle does not allow the optimal $\gamma$ value to be selected) it is noticeable that the
full adjoint  method is able  to retrieve  a small quantity  of information, along  with a
large  noise (optimal control  problem obviously  ill-posed) whereas  the ``self-adjoint''
method does not provide anything else than the initial guess, irrespective of the value of
$\gamma$.\\

\section{Density of the data}
\label{sec54}

The previous simulations  have been performed using surface velocity  data quite dense (one
measurement  every $dx$).   This section  deals with test  cases
identical  to  the previous
section but  using sparser (one  measure point every  $1$km) and thus more  realistic data
(corresponding  to  one ice  thickness,  see  \eg  \citet{gudmundsson2008}).  This  density
corresponds to  approximately $10$  times less  measurement points than  the previous  case. We
consider thereafter the following friction coefficient for the synthetic data:

\beq\label{fric4f3}
\beta_r^N(x)=a+\frac{a}{2}\sin\left(\frac{2\pi
    x}{200dx}\right)+\frac{a}{5}\sum_{i=1}^N f_i(x) \eeq with

\beq
f_i(x)=\sin\left(\frac{2\pi x}{w_i dx}\right) \mbox{ with } w_1=100,~w_2=50,~w_3=20,
\eeq

 and by extension, we set: 

\beq
\bs{f_0}(x)=\sin\left(\frac{2\pi x}{w_0 dx}\right) \mbox{ with } w_0=200.
\eeq

\begin{figure*}[t] \centering
\includegraphics[scale=0.9]{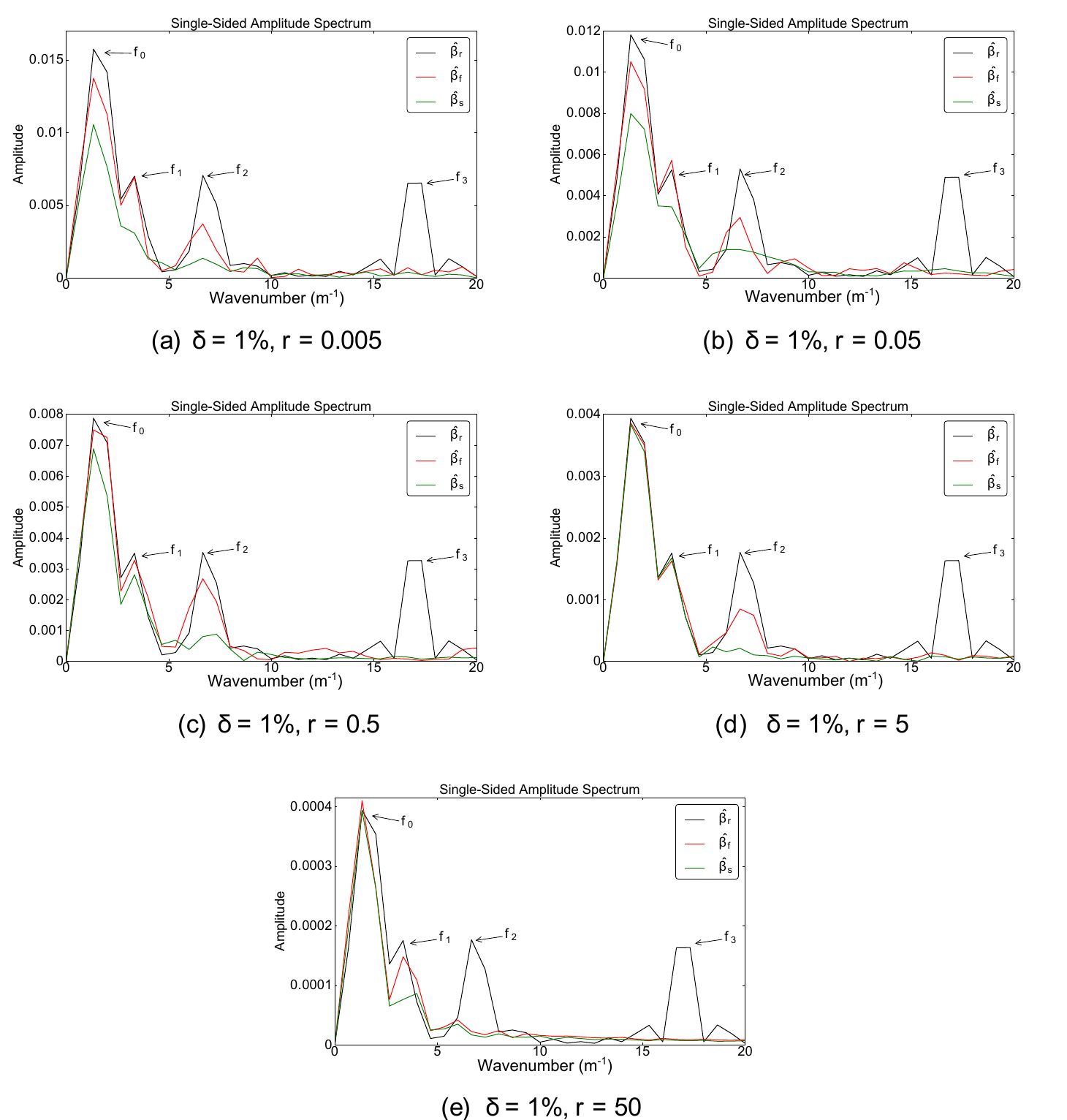}
 \caption {Discrete Fourier  Transform for inferred  friction coefficients $\beta_f$  and $\beta_s$
   and for the target one $\beta_r$ for sparse data a $1\%$ noise level.}
 \label{fftmertzmask}
 \end{figure*}

The friction coefficient  chosen for these simulation contains  lower frequencies than the
previous  one,  simulating  a  carrier  wave  of  wavelength  $20h$  perturbed  by  high
frequencies of  wavelengths $10h$, $5h$ and  $2h$. These characteristics  are summarized in
Table \ref{params3}. Results are plotted in  Figure \ref{fftmertzmask} for a noise level of
$1\%$.

\begin{table*}
\begin{center}
\begin{tabularx}{0.7\textwidth}{X||c|c|c|c} & $\bs{f_0}$ &
$f_1$ & $f_2$ & $f_3$ \\ 
\hline\hline 
Wavelength w.r.t $h=1$km (thickness) & $20h$ & $10h$ & $5h$ & $2h$ \\ 
\hline
Wavelength w.r.t $dx=100$m (edge length) & $200dx$ & $100dx$ & $50dx$ & $20dx$ \\
\hline
Wavenumber w.r.t $L=33.3$km (domain length) & $1.66$m\textsuperscript{-1} & $3.33$m\textsuperscript{-1} & $6.66$m\textsuperscript{-1} & $16.6$m\textsuperscript{-1}
\end{tabularx}
\end{center}
\caption{Characteristics of signal $\beta$ given by \eqref{fric4f3}.}\label{params3}
\end{table*}

 As a  consequence, the level of identifiability  assessed for dense data  in the previous
 section is  no longer valid.  However,  considering that one  out of ten points  has been
 retained, results seem rather convincing. The  full adjoint method is able to accurately recover
 frequencies of  wavelengths $20h$ and  $10h$ (corresponding to  $f_0$ and $f_1$)  for all
 degrees  of  slip. The  ``self-adjoint''  method recovers  the  carrier  wave quite  well
 although a stronger friction ($r\leq  0.05$) significantly degrades the reconstruction of
 the amplitude.   Frequency $f_1$ is well  captured for propitious  situations ($0.5\leq r
 \leq 5$).   Frequency $f_2$ (of wavelength  $5h$, the lowest frequency  considered in the
 dense data  situation) is partially reconstructed by  the full
 adjoint method for  $r\leq 5$ and
 never captured by the ``self-adjoint'' method.

The case $r=50$ is a lot less problematic than previously found, due to lower frequencies and
subsequently less local effects regarding the sharpness of the bed discretization. A pronounced
difficulty   appears  for   the  identification   of  frequency   $f_1$   (of  wavelength
$10h$). The  case $r=50$  is the only  one where  frequency $f_2$ does  not appear  in the
spectrum of $\widehat{\beta}_f$ (consistently to the previous simulations).

\conclusions

The  significant time  saving brought  by  the ``self-adjoint''
method due to its straightforward implementation is a favorable asset.
 However, its  reliability is  questionable and  it seems
important to know its limitations in order to perform realistic
experiments.

The  realistic simulation  (low density  data,  $1\%$ noise,  real topography,  non-linear
friction) allows us to assess the full adjoint method ability to accurately identify
wavelengths  greater or equal  to $10$ ice  thicknesses and to capture  effects of
wavelengths up to $5$ thicknesses for slip  ratio lower than $5$. These bounds are defined
by the level of noise considered on the data and a higher accuracy on the data would allow
to identify higher frequencies.\\

The ``self-adjoint'' method , based on  second order numerical schemes, while providing an
incorrect gradient, is  able to reconstruct wavelengths greater  than $20$ ice thicknesses
(with noticeable  difficulties for strong  friction). Wavelengths of $10$  ice thicknesses
can be  captured in propitious  situations of intermediate  sliding ($0.5\leq r  \leq 5$).
These  bounds are  strict  and a  lower  noise would  not allow  to  overcome the  limited
precision of the ``self-adjoint'' gradient.\\

The results provided by the full  adjoint method are significantly better than those given
by  \citet{Petra2012}  (who assess  a  limit  of $20$  ice  thicknesses  for a  non-linear
rheology). It is  difficult to compare considering that the  authors provide neither their
slip ratio nor the density of the  data. In addition, the authors of \citet{Petra2012} consider
a linear  friction law.\\

The  use  of  a non-linear  friction  allows  us  to  simulate  complex behaviors  of  the
ice-bedrock interaction.  This type of  law can describe  a non-linear deformation  of the
basal substrate or a non-linear response of  the sliding velocity to the water pressure of
sub-glacial cavities.  The former reconstructions focus on the identification of a generic
$\beta$. However one may confidently generalize these results to more complex sliding laws
where $\beta$  would be identified  through its parameterization  (by a water  pressure, a
contact surface  with sub-glacial  cavities, a sedimentary  roughness, a  geothermal flux,
...). It  is important to  recall that an  identical cost $j$  does not mean  an identical
results due  to the equifinality aspect  and that over-parameterization is  hardly ever in
favor   of  an  accurate   identification;  the   identification  of   several  parameters
simultaneously would strongly reinforce the problem of equifinality (\ie the ill-posedness
of the inverse problem).  

In other respect, we recall that the use of a ``self-adjoint'' gradient
in the  case of  a non-linear friction  law leads  to ignore important  extra contribution
(compared to the case of a linear friction, see Section \ref{adjeq}) in the gradient computation (see
equation \eqref{fric}) which adds even more discrepancy into the adjoint problem.\\

This work focuses on the identification of  the friction coefficient that plays a major role
to  control the  flow (\ie  the model  shows  a great  sensitivity to  the friction).  The
identification  of a  parameter  such as  the  consistency $\eta_0$,  for  which the  model
sensitivity is significantly lower, needs to be done with caution for the full adjoint
method (see \eg \citet{iopip2014}) and thus with increased caution for the ``self-adjoint'' method.

Finally, the adjoint obtained  from source-to-source algorithmic differentiation allows us to
simulate every level  of needed precision between the best precision  of the exact adjoint
to the lowest one of the ``self-adjoint'' approximation.  This leads to the consideration of
an  \textit{incomplete   adjoint}  methodology  where  the   approximation  is  completely
adjustable,  thus  allowing the  right  compromise  between  CPU-time, memory  burden  and
required accuracy to be achieved.  Numerical experiments show that the retention of
the last two  states of the forward iterative loop (or equivalently
the first two states of the reverse accumulation loop) within the gradient computation  significantly improves its precision
while maintaining a quite small computational burden. Let us recall that such an 
approach should be combine with an accurate solution for the forward problem\\

\begin{acknowledgements}
  This work  was partially  supported by  PRES Toulouse, with  the PhD  fund of  the first
  author. The authors want to thank Benoit  Legrésy (LEGOS) for his topography data on the
  Mertz glacier and Ronan Madec (IMT) for his help on the development of the adjoint model
  and  the writing  of appendix  \ref{annexe}. The  authors also  want to  thank  the four
  reviewers for  their careful  and sound reviews  leading to  an improved version  of the
  present work.This  work was also supported  by Agence Nationale de  la Recherche through
  ADAGe project No.  ANR-09-SYSC-001.
\end{acknowledgements}

\bibliography{biblio}
\FloatBarrier
\appendix

\section{Adjoint of a linear solver}
\label{annexe}
This appendix describes how to generate the adjoint of a generic routine
containing a call to a linear solver whose contents is a priori unknown.\\ 

\paragraph*{The direct routine}

\begin{figure}
\begin{center}
\includegraphics[width=.9\hsize]{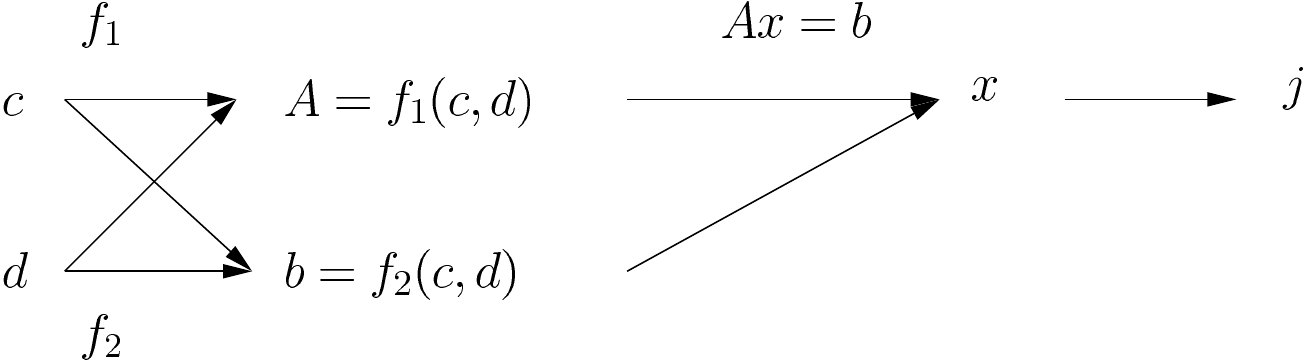}
\caption{Direct routine scheme}
\label{fig:directroutine}
\end{center}
\end{figure}

A general direct routine can be described as follows.
Let $c$  and $d$ be two given input parameters such that:
$$
\left( \begin{array}{c} A \\ b \end{array} \right) = f(c,d) = \left( \begin{array}{c} f_1(c,d) \\
f_2(c,d) \end{array} \right)
$$
with  $A$ a  matrix and  $b$ a  vector.  Let  $x$ be  the solution  of the  linear system
$Ax=b$ and $j$ defining a cost function evaluated at $x$. Figure \ref{fig:directroutine} illustrates
the direct routines dependencies.\\

\paragraph*{The linear tangent routine}
\label{sub:lintanrout}

The \textit{linear
  tangent routine} associated to the direct routine described
herebefore is then written:\\
$$
 \left( \begin{array}{c} \dot A \\ \dot b \end{array} \right) = df(c,d) \cdot \left( \begin{array}{c} \dot c
\\ \dot d \end{array} \right)
$$\\
where $df$ is the linear tangent model and $\dot c$ and $\dot d$ are
the tangent variables corresponding to  parameters $c$ and
  $d$. They serve as input parameters for the linear tangent model in
  order to compute $\dot  A$ and $\dot b$. We can now differentiate
  the linear system operation $Ax=b$ to obtain the following \textit{linear
  tangent system}:

$$ A\dot x = \dot b - \dot A x .$$

\begin{figure}[h!]
\begin{center}
\includegraphics[width=.9\hsize]{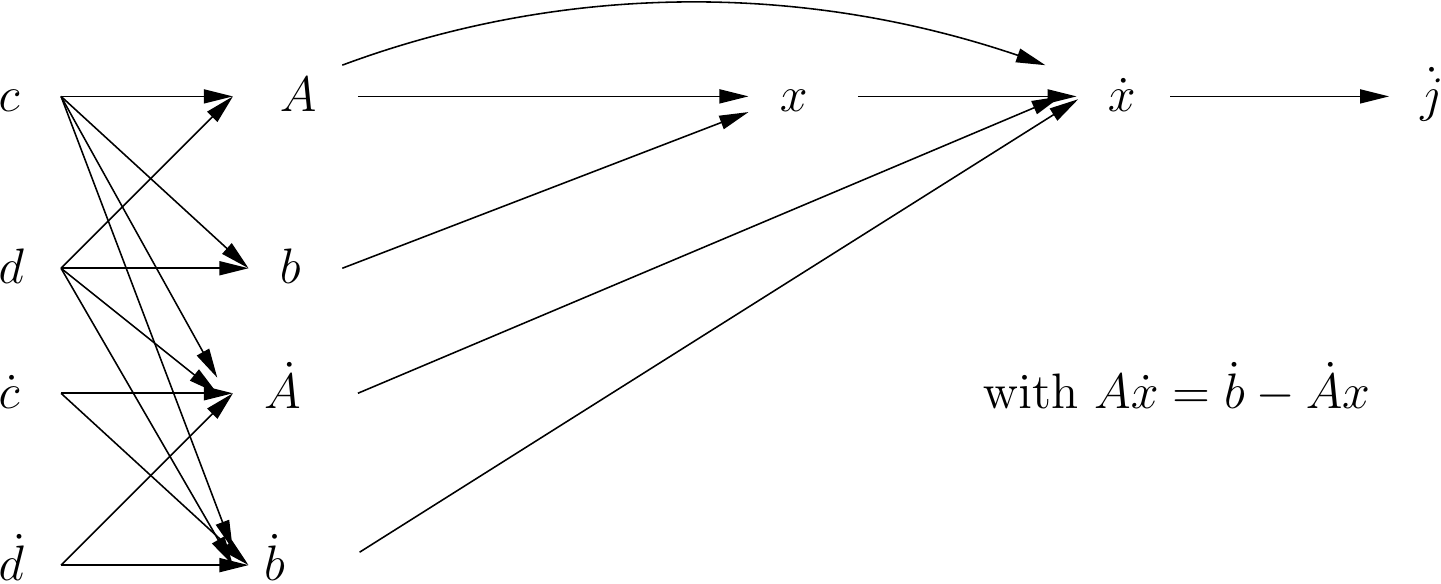}
\caption{Linear tangent routine scheme}
\label{fig:lintangentroutine}
\end{center}
\end{figure}

The matrix $A$ and  the vector $x$ are provided by the direct  routine
and the quantities $\dot A$ and $\dot b$ are given by the  tangent
linear routine. The linear solver is finally called, as a black-box,
to solve this equation and to obtain the linear tangent unknown
$\dot x$ where the gradient of the cost $\dot j$ can be evaluated.  The quantity
$\dot x$ represents the derivative value of  $x$
at  $(c,d)$ in a given  direction  $(\dot  c,  \dot  d)$. The linear
tangent routine is illustrated in Figure \ref{fig:lintangentroutine}.\\

\paragraph*{The generated adjoint routines}
\label{sub:adjointrout} 

Let us recall that the adjoint code  corresponds to the linear tangent
code in a reverse order.  It follows that the output variables of the
the linear tangent routine are input variables for the adjoint
routine. Therefore, the output  variables  of the  adjoint  routine are  $\bar c$  and  $\bar d$  and
represent the adjoint  variables of $(c,d)$ (and are consequently of  same type and size).
The  adjoint cost  $\bar  j$  is the  input  variable of  the  adjoint  cost function  and
similarly, the adjoint state $\bar x$ is the input variable of the adjoint linear system.\\

The computation of the adjoint state can be split into three steps (see Fig. \ref{fig:lintangentroutine}):
\begin{enumerate}
\item From $\bar j$, obtain $\bar x$ (generally provided by an
  independent routine called the adjoint cost function)
\item From $\bar x$, obtain $\bar A$ and $\bar b$
\item From $\bar A$ and $\bar b$, obtain $\bar c$ and $\bar d$
  using the adjoint model $df^*$ such that:
$$ \left( \begin{array}{c} \bar c \\ \bar d \end{array} \right) = df^*(c,d) \cdot
\left( \begin{array}{c} \bar A \\ \bar b \end{array} \right).
\label{eq.adjoint.step3}
$$
\end{enumerate}


\paragraph*{The adjoint of the linear system}
\label{sub.lin.sys} The linear solver call occurs in the second
step. The input variable  is $\bar x$ and the  output variables are $\bar A$  and $\bar b$.
In the linear tangent code, we have: $A\dot x = \dot b - \dot A x$ or
if one splits it into two steps:
$$\begin{array}{lcl}
2a. &\dot {b'}  & = \dot b - \dot A x, \\ 
2b. & A\dot x & = \dot {b'}.
\end{array} $$

An adjoint calculation being performed in the reverse order,  the
adjoint of this procedure starts with instruction $2b$ which can be written as follows:
$$ \left( \begin{array}{c} \dot x \\ \dot b' \end{array} \right) = \left( \begin{array}{ccc}
0&A^{-1} \\ 0&1
\end{array} \right) \times \left( \begin{array}{c} \dot x \\ \dot b'
\end{array} \right) \, .
$$
Since $ \dot {b'} $ is the input variable for the instruction $2b$,
its adjoint counterpart $ \bar b'$ is the output of the adjoint
instruction of $2b$ (by convention, the adjoint output variables are set to
$0$ before entering the adjoint routine). Similarly, since $ \dot {x}$
is  the output  variable, its  adjoint counterpart $ \bar x$ is  an
input one.   The adjoint instruction of step $2b$ are then written:
$$
\left( \begin{array}{c} \bar x \\ \bar b' \end{array} \right) = \left( \begin{array}{ccc} 0&0 \\ A^{-T} &1
\end{array} \right) \times \left( \begin{array}{c} \bar x \\ \bar b'
\end{array} \right) \, .
$$

The variable $ \bar b'$ is an output variable, hence set to $0$ before
entering the adjoint routine. This operation corresponds
to solving the linear system $A^{T}
\bar b'= \bar  x $ in order to obtain $\bar b'$ (using the
linear solver).\\

Once $\bar b'$ has been computed, one has to perform the adjoint of
the instruction $2a)$. This instruction can be written as the
following linear operation:
$$
  (\dot b' , \dot b, \dot A) = (\dot b' , \dot b, \dot A) \times \left( \begin{array}{ccc} 0&0&0 \\ 1&1&0 \\
-x&0&1
\end{array} \right).
$$

The corresponding adjoint instruction is written:
$$\displaystyle 
(\bar b' , \bar b, \bar A) = (\bar b' , \bar b, \bar A) \times \left( \begin{array}{ccc} 0&1&-x^T \\0&1&0 \\
0&0&1
\end{array} \right)
$$

which leads, in reverse order, to perform the following operations:
$$
\left\{ \begin{array}{rcl} \bar b &= & \bar b' ,\\ \bar A &= &- \bar b x^T .
\\
\end{array} \right.
$$

The variables $\bar A, ~ \bar b$ are output variables, hence set to $0$ before entering the adjoint
routine and $ \bar b'$ has been obtained from the previous step (the
adjoint of step $2b$).\\

In summary, the adjoint of the tangent linear instructions $A\dot x =
\dot b - \dot A x$ (referred as step 2)) can be written:
$$
\left\{ \begin{array}{rcl} A^{T} \bar b & = & \bar x ,\\ \bar A &= &-\bar b x^T ,\\ \bar x &=& 0 .\\
\end{array} \right.
$$
where $(\bar c, \bar d)$ are the components of the gradient with
respect to $(c,d)$ obtained from the adjoint model $df^*$.
The first instruction  can then be solved using the  same linear solver as the one  used in the
direct routine. The second instruction is written: $\bar A_{ij} = - \bar b_i x_j$.\\

Let us point out that the matrix $\bar A$ is of the same type as $A$ with the same sparse
profile (even if $-\bar b x^T $ is a priori a full matrix).  Therefore, only the
adjoint values of coefficients $A_{i,j}$ are required.\\
 
\begin{figure}[h!]
\begin{center}
\includegraphics[width=.9\hsize]{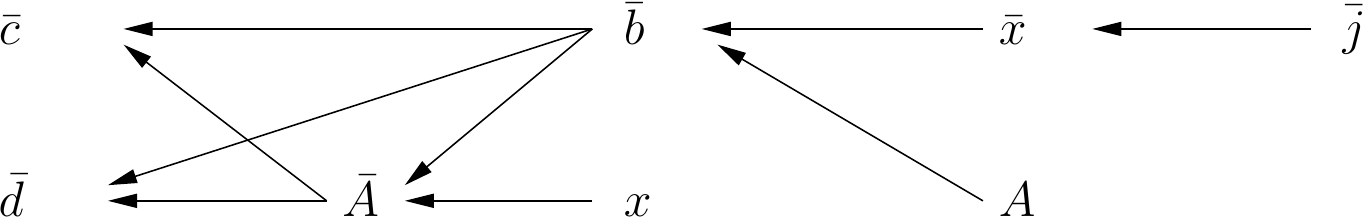}
\caption{Adjoint routine representation}
\label{fig:adjointroutine}
\end{center}
\end{figure}

The steps of the adjoint routine are illustrated in Figure \ref{fig:adjointroutine}.\\

\end{document}